\documentclass[pre,aps,reprint]{revtex4-2}

\usepackage{amsmath}          
\usepackage{amstext}          
\usepackage{amssymb}          
\usepackage{amsfonts}         
\usepackage{graphicx}         
\usepackage{pict2e}           
\usepackage{color}            
\usepackage{hyperref}

\newcommand{\Var}{\operatorname{Var}}

\newcommand{\dd}{\mathrm{d}}

\newcommand{\ii}{\mathrm{i}}

\begin{document}
\title{Cheaper access to universal fluctuations in integrable spin chains\\from boundary effects}
\author{Sylvain Prolhac}
\email{sylvain.prolhac@irsamc.ups-tlse.fr}
\affiliation{Laboratoire de Physique Th\'eorique, Universit\'e de Toulouse, France}

\begin{abstract}
	Observing super-diffusive fluctuations from Kardar-Parisi-Zhang (KPZ) universality in isotropic integrable spin chains is usually challenging as it requires a fairly large number of spins in interaction. We demonstrate in this paper, in the context of classical spins, that accounting for boundary effects lowers the bar, down to a few dozen spins in some cases. Additionally, boundaries control the relaxation to stationarity, which leads to many new universal scaling functions to explore, both in periodic spin chains and for open chains with magnetization imposed by reservoirs at the ends.
\end{abstract}

\maketitle

One-dimensional integrable many body systems, which possess extensively many local conserved quantities and feature unusual transport properties, have been the subject of intense theoretical progress in the past decade \cite{BCDNF2016.1,CADY2016.1,BHMKPSZ2021.1,S2024.1}. In the context of classical and quantum integrable spin chains, super-diffusion was in particular observed for spin transport in chains with isotropic interaction, both numerically \cite{Z2011.1,LZP2017.1} and experimentally \cite{SSDNSGMT2021.1,WRYMKSHRGYBZ2022.1,KRMZG2023.1,R+2024.1}.

Super-diffusion in integrable spin chains has been related \cite{LZP2019.1,DKSD2019.1,DNMKI2019.1,DM2020.1,KP2020.1,WSBE2020.1,BGI2021.1,YMKHY2022.1,RDSK2023.1,KSIP2024.1,TTBFVDN2025.1,PMDSVPZ2025.1,GV2024.1} to Kardar-Parisi-Zhang (KPZ) fluctuations, initially introduced in the study of interface growth \cite{KPZ1986.1} and driven particles \cite{vBKS1985.1}, and which generically emerge in systems with few conservation laws from non-linear fluctuating hydrodynamics \cite{vB2012.1,S2014.1}. In integrable spin chains, KPZ fluctuations for spin transport were argued to result from a decoupling of magnetization from the infinitely many conservation laws \cite{B2020.1}, and a description in terms of two KPZ modes \cite{DNGV2023.1} reproduces the spin flip symmetry of the chain from KPZ fluctuations, which are inherently asymmetric due to their directed nature. Recent works \cite{R+2024.1,KSIP2024.1,TTBFVDN2025.1,MSBDZX2025.1} on higher correlation functions have however cast some doubt on this simple picture.

So far, most research on KPZ fluctuations \footnote{More specifically, the KPZ fixed point, corresponding to a high non-linearity limit of the KPZ equation \cite{KPZ1986.1}} in relation with integrable spin chains has focused on the idealized \emph{infinite line} setting \cite{C2011.1,HHT2015.1,S2017.1,T2018.1}, where the system is large enough so that on the time scales considered, boundaries have no influence. There, KPZ fluctuations describe a scale invariant random height field $h(x,t)$ depending on space $x\in\mathbb{R}$ and time $t\geq0$, whose standard deviation grows as $t^{1/3}$ and correlation length as $t^{2/3}$. Then, in a given physical system of total size $L$ with microscopic scale $\ell$ fixed by e.g. lattice spacing, KPZ fluctuations on the infinite line may only be observed in the regime $\ell\ll T^{2/3}\ll L$ with $T$ the microscopic time scale.

Lifting the restriction $T^{2/3}\ll L$ and considering rather the relaxation scale with finite KPZ time $t\sim T/L^{3/2}$, boundary effects begin to be felt and long range correlations eventually span the whole system: this is the regime of KPZ fluctuations \emph{in finite volume} \cite{P2024.1}, describing the crossover between KPZ on the infinite line at short time $t$ and stationary fluctuations at late time $t$, when the initial condition is eventually forgotten. This regime was considered recently in one-dimensional condensates \cite{ACC2024.1} and stationary spin chains \cite{RDSK2024.1}. Crucially, correlation functions depend on boundary conditions in this regime, and we consider in this letter periodic boundaries $x\equiv x+1$ (setting by convention the system size to $1$ in the following), and open boundaries with fixed boundary values $\sigma(0,t)=\sigma_{a}$, $\sigma(1,t)=\sigma_{b}$ for the conserved density field $\sigma=\partial_{x}h$ \footnote{Since the height field is typically nowhere differentiable, some care is needed to define the density field, and the boundary values $\sigma_{a}$, $\sigma_{b}$ may be understood in practice by taking the spatial derivative after averaging.}, corresponding for the spin chain to fixed boundary magnetization at the ends, as explained below.

\begin{figure}[b]
	\includegraphics[width=86mm]{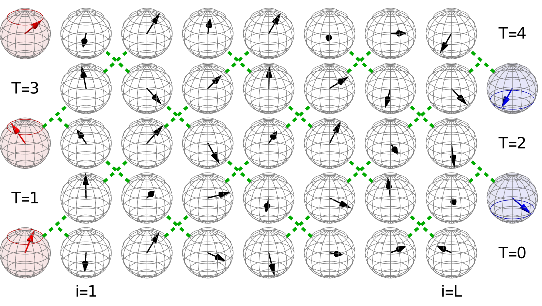}
	\vspace{-5mm}
	\caption{Graphical representation of the dynamics of the KPLL model with open boundaries. Spins at various sites $i$ and times $T$ are coupled according to the dashed green links.}
	\label{fig KPLL}
\end{figure}

The aim of this letter is to demonstrate, in the specific setting of classical integrable spin chains, that accounting for boundary effects leads to an easier access to KPZ fluctuations, by lowering the number of spins needed to reach the universal regime, especially with periodic boundaries, and by providing many new scaling functions to test. Additionally, stationary fluctuations such as the ones observed in this letter for open chains with fixed magnetization imposed by reservoirs at the boundaries are usually more tractable analytically \cite{P2015.4} than the full dynamics, which potentially opens the way to a proper derivation of KPZ fluctuations for spin chains.

Two classes of correlation functions have been considered in spin chains in relation with KPZ, involving either the local magnetization or the associated current. In the former case a full agreement has been observed on the infinite line with corresponding correlations of a single KPZ density field $\sigma$. In the latter case, correlations of a symmetric magnetization current only match partially those of a single asymmetric KPZ height $h$. An explanation in terms of two independent KPZ heights with opposite asymmetry \cite{DNGV2023.1} appears to be at odds with recent experimental and numerical works \cite{R+2024.1,KSIP2024.1,TTBFVDN2025.1,MSBDZX2025.1}. In this letter, we focus for simplicity on correlations of the local magnetization only, on the time scale where KPZ fluctuations relax to their stationary state. With periodic boundaries, correlations in the spin chain match well with those of a single KPZ density field. With open boundaries, however, we find that two KPZ fields are needed.

The integrable spin chain studied in this letter, known as the Krajnik-Prosen-Landau-Lifshitz (KPLL) model \cite{KP2020.1}, features $L$ classical spins taken as 3-dimensional unit vectors $\mathbf{S}_{i}(T)$ evolving in discrete time $T=0,1,2,\ldots$ by simultaneous action of a map $\varphi_{\tau}$ on all pairs of spins $(\mathbf{S}_{2j-1},\mathbf{S}_{2j})$ at even time steps and $(\mathbf{S}_{2j},\mathbf{S}_{2j+1})$ at odd time steps
as $(\mathbf{S}_{i},\mathbf{S}_{i+1})\mapsto(\varphi_{\tau}(\mathbf{S}_{i},\mathbf{S}_{i+1}),\varphi_{\tau}(\mathbf{S}_{i+1},\mathbf{S}_{i}))$, which preserves $\mathbf{S}_{i}+\mathbf{S}_{i+1}$, with
\begin{equation}
	\label{local map KPLL}
	\varphi_{\tau}(\mathbf{S},\mathbf{S}')=\frac{a\,\mathbf{S}+\tau^{2}\,\mathbf{S}'+\tau\,\mathbf{S}\wedge\mathbf{S}'}{a+\tau^{2}}\;.
\end{equation}

The coefficient $a=(1+\mathbf{S}_{1}\cdot\mathbf{S}_{2})/2$ ensures that the norm of each spin is conserved. Additionally, in order to smooth out the staggered microscopic structure resulting from the two-step nature of the dynamics, data are always averaged over two consecutive time steps.

The parameter $\tau$ of the KPLL model has to be chosen with some care. For evaluating the stationary magnetization profile, we use $\tau=1$ for fast convergence to the stationary state. For the stationary two-point correlation function, smaller values of $\tau$ appear to make convergence with $L$ faster at the price of slowing down the effective dynamics, and we found that $\tau=0.1$ was a good compromise for the values of $L$ considered in this letter.

For open boundaries, we couple the system to reservoirs fixing the magnetization on the left to $S_{a}$ (on the right to $S_{b}$) at even (odd) time steps, modelled by adding at the appropriate boundary a spin $\mathbf{S}$ randomly chosen at each time step uniformly on the circle $S^{z}=S_{a}$ ($S^{z}=S_{b}$), see figure~\ref{fig KPLL}. The large scale dynamics depends on the amplitude of boundary magnetizations with respect to the system size $L$. When $S_{a}$ and $S_{b}$ are of order one, oscillations of period $\sim L^{2}$ decaying on the time scale $T\sim L^{3}$ are present. On the other hand, the natural regime for KPZ fluctuations in finite volume corresponds to boundary magnetizations scaled down to the amplitude of typical fluctuations within the chain, as
\begin{equation}
	\label{Sab[sab]}
	S_{a}=\frac{s_{a}}{\sqrt{L}}
	\quad\text{and}\quad
	S_{b}=\frac{s_{b}}{\sqrt{L}}\;.
\end{equation}

Away from boundaries, the dynamics of the KPLL model depends on a single parameter $\tau>0$ which for small $\tau$ merely fixes the time scale. Then, in the limit $\tau\to0$, the KPLL model converges after rescaling $T$ by $\tau$ to the continuous time Ishimori chain \cite{I1982.1}, which we consider only with periodic boundaries in the following, and is defined by
\begin{equation}
	\frac{\dd\mathbf{S}_{i}}{\dd t}=\frac{\mathbf{S}_{i}\wedge\mathbf{S}_{i+1}}{1+\mathbf{S}_{i}\cdot\mathbf{S}_{i+1}}-\frac{\mathbf{S}_{i-1}\wedge\mathbf{S}_{i}}{1+\mathbf{S}_{i-1}\cdot\mathbf{S}_{i}}\;.
\end{equation}

Randomness is generated solely by the initial condition (and also the boundary spins for the open chain), since the KPLL dynamics is otherwise fully deterministic. In order to observe stationary KPZ fluctuations, we focus for periodic boundaries on spin chains in thermal equilibrium at infinite temperature, corresponding to independent spins drawn uniformly on the sphere $||\mathbf{S}_{i}||=1$. For open boundaries, this uniform state is no longer stationary, but a unique nonequilibrium stationary state is reached at late times, which we choose as our initial condition. Stationary averages for the spin chain are denoted by the subscript $_{\rm st}$ in the following.

For isotropic integrable spin chains in a stationary state, super-diffusive fluctuations are observed in the local magnetization in e.g. the direction $\rm z$, defined as
\begin{equation}
	\label{m[S]}
	m(x,t)=\frac{S_{i}^{\rm z}(T)-R}{\sqrt{\Var(M)_{\rm st}}}
\end{equation}
with $i=xL$, and $T\sim tL^{3/2}$ (see discussion below). For periodic boundaries, the shift $R$ must be equal to the conserved total magnetization
\begin{equation}
	\label{M}
	M=\frac{1}{L}\sum_{i=1}^{L}\mathbf{S}_{i}^{z}\;,
\end{equation}
such that $\Var(M)_{\rm st}\sim L^{-1}$, in order to enforce periodicity for $\int_{0}^{x}\dd y\,m(y,t)$. We set $R=0$ for open boundaries.

Because of the global spin flip symmetry $\mathbf{S}_{i}\leftrightarrow-\mathbf{S}_{i}$ of the dynamics, the local magnetization (\ref{m[S]}) can not in general be identified with a single KPZ density field $\sigma=\partial_{x}h$ but rather with the sum of two KPZ densities with opposite asymmetry, which we assume to be \emph{independent}.

For periodic boundaries, the KPZ density field is actually symmetric (unlike the height), and writing $m(x,t)\simeq\sigma(x,t)$ allows to compute correlation functions of the magnetization at large $L$ from a single KPZ field in that case. We emphasize that the correspondence between the magnetization $m(x,t)$ for the spin chain and the KPZ density field $\sigma$ does not require an extra normalization for periodic boundaries since for a stationary state constituted of independent spins distributed uniformly on the sphere, the quantity $\int_{0}^{x}\dd y\,m_{\rm st}(y)$ converges at large $L$ to a standard Brownian bridge, which is the stationary state for the KPZ height.

For open boundaries on the other hand, the KPZ density fields $\sigma^{\sigma_{a},\sigma_{b}}$ and $-\sigma^{-\sigma_{a},-\sigma_{b}}$, of opposite asymmetry, have in general completely unrelated probability distributions, and the natural hypothesis with two KPZ fields restoring spin flip symmetry is simply
\begin{equation}
	\label{m[sigma_a,sigma_b]}
	m(x,t)\simeq\frac{\sigma^{\sigma_{a},\sigma_{b}}(x,t)-\sigma^{-\sigma_{a},-\sigma_{b}}(x,t)}{\mathcal{N}(\sigma_{a},\sigma_{b})}\;.
\end{equation}
The normalization $\mathcal{N}(\sigma_{a},\sigma_{b})=\sqrt{\Var(h_{\rm st}(1))}$ where $h_{\rm st}=h_{\rm st}^{+}+h_{\rm st}^{-}$ is built from independent stationary KPZ heights $h_{\rm st}^{\pm}(x)=\pm\int_{0}^{x}\dd y\,\sigma_{\rm st}^{\pm\sigma_{a},\pm\sigma_{b}}(y)$. An explicit expression is available for $\mathcal{N}(\sigma_{a},\sigma_{b})$, see supplemental material (SM) \cite{SM}.

\begin{figure}
	\includegraphics[width=42mm]{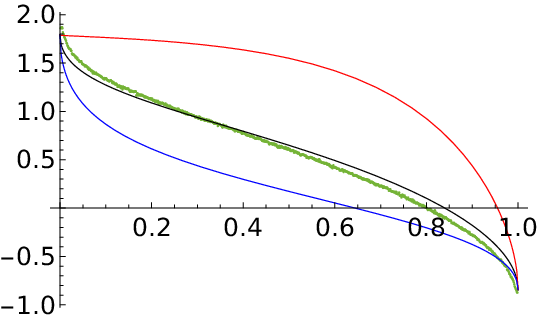}
	\includegraphics[width=42mm]{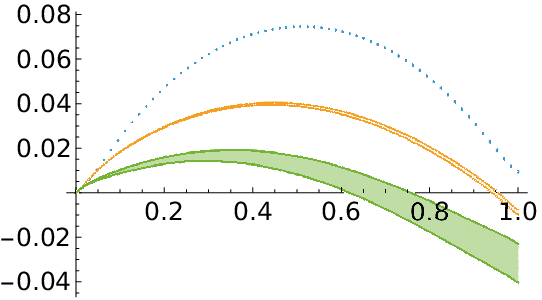}
	\begin{picture}(0,0)
		\put(-52,21.5){(a)}
		\put(-47,10){$x$}
		\put(-66,21.5){$C_{1}(x)$}
		\put(-7,21.5){(b)}
		\put(-3.5,10){$x$}
	\end{picture}\\
	\includegraphics[width=42mm]{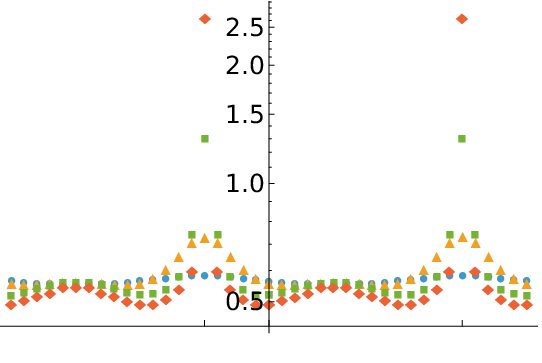}
	\hspace*{10mm}
	\includegraphics[width=30mm]{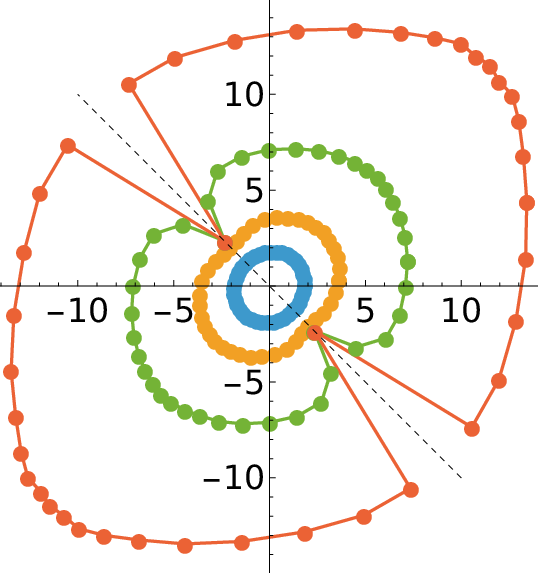}
	\begin{picture}(0,0)
		\put(-85,25){(c)}
		\put(-62,21){$L^{1/2}\Var(M)$}
		\put(-74,0){$-\pi/4$}
		\put(-53,0){$3\pi/4$}
		\put(-87,15){
			\begin{tabular}{rlc}
				$|\mathsf{s}|=1$ &\hspace*{-2.7mm}& \begin{tabular}{c}\includegraphics[width=1.25mm]{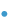}\end{tabular}\\[-1mm]
				$2$ &\hspace*{-2.7mm}& \begin{tabular}{c}\includegraphics[width=1mm]{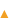}\end{tabular}\\[-1mm]
				$4$ &\hspace*{-2.7mm}& \begin{tabular}{c}\includegraphics[width=0.8mm]{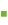}\end{tabular}\\[-1mm]
				$8$ &\hspace*{-2.7mm}& \begin{tabular}{c}\includegraphics[width=1.25mm]{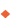}\end{tabular}
			\end{tabular}
		}
		\put(-33,25){(d)}
		\put(-3.7,14.3){$m_{a}$}
		\put(-20.5,30.5){$m_{b}$}
		\put(-12.7,10){\rotatebox{-45}{$s_{a}+s_{b}=0$}}
	\end{picture}
	\caption{(a) Stationary magnetization profile $C_{1}(x)$ plotted as a function of the position $x=i/L$. The green dots represent results from numerical simulations of the KPLL model with $L=1023$ spins, $T=10^{5}$ time steps, averaged over $10^{7}$ realizations, with boundary parameters $s_{a}=1$, $s_{b}=-0.5$. The solid black line in the middle is the exact KPZ result (\ref{C1[c1-c1] open}) normalized accordingly, with boundary densities $\sigma_{a}\approx1.33$, $\sigma_{b}\approx-0.63$ adjusted from $\langle S_{1}^{\rm z}(T)\rangle$, $\langle S_{L}^{\rm z}(T)\rangle$. The solid red and blue lines below and above correspond to keeping a single KPZ mode. (b) Difference between the KPLL simulations and the exact result (statistical average $\pm$ one standard deviation) for the stationary height profile (integral of $C_{1}(x)$ with respect to $x$), with $L=63,255,1023$ spins from top to bottom, roughly compatible with finite size corrections vanishing as $L^{-1/2}$. (c) Variance of the total magnetization $L^{1/2}\Var(M)$ with $L=511$ spins plotted as a function of $\arg\mathsf{s}$ for $\mathsf{s}=s_{a}+\ii s_{b}$. (d) $m_{a},m_{b}$ from (\ref{eq boundaries}) for the values of $s_{a},s_{b}$ in (c).}
	\label{fig C1}
\end{figure}
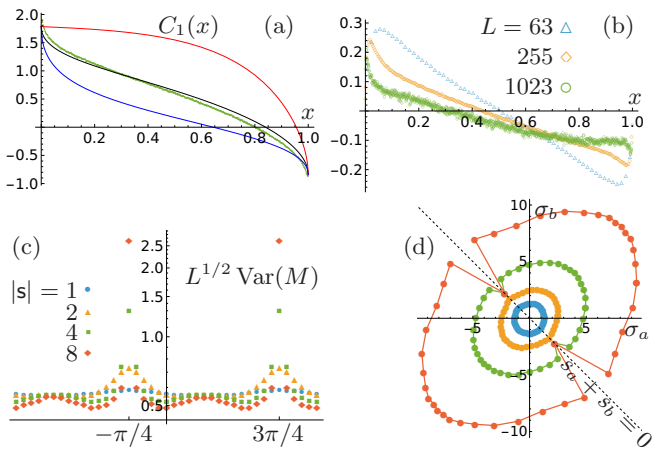

The hypothesis (\ref{m[sigma_a,sigma_b]}) implies that $\langle m(0,t)\rangle\simeq m_{a}(\sigma_{a},\sigma_{b})$ and $\langle m(1,t)\rangle\simeq m_{b}(\sigma_{a},\sigma_{b})$ with $m_{a}=2\sigma_{a}/\mathcal{N}(\sigma_{a},\sigma_{b})$ and $m_{b}=2\sigma_{b}/\mathcal{N}(\sigma_{a},\sigma_{b})$. The KPZ boundary densities $\sigma_{a}$, $\sigma_{b}$ can then be extracted from the simulations of the KPLL model by solving numerically for $\sigma_{a}$ and $\sigma_{b}$ the system of two equations
\begin{equation}
	\label{eq boundaries}
	\frac{\langle S_{1}^{\rm z}\rangle_{\rm st}^{s_{a},s_{b}}}{\Var(M)_{\rm st}^{1/2}}=m_{a}
	\quad\text{and}\quad
	\frac{\langle S_{L}^{\rm z}\rangle_{\rm st}^{s_{a},s_{b}}}{\Var(M)_{\rm st}^{1/2}}=m_{b}\;,
\end{equation}
which leads in particular when $s_{a}+s_{b}=0$ to the prediction that $\langle S_{1}^{\rm z}\rangle_{\rm st}^{s_{a},s_{b}}/\Var(M)_{\rm st}^{1/2}$ stays bounded by $2\sqrt{3}$ in the limit of a large number $L$ of spins, associated with a walnut shape for the map $(s_{a},s_{b})\mapsto(m_{a},m_{b})$, see figure~\ref{fig C1}d and SM \cite{SM}.

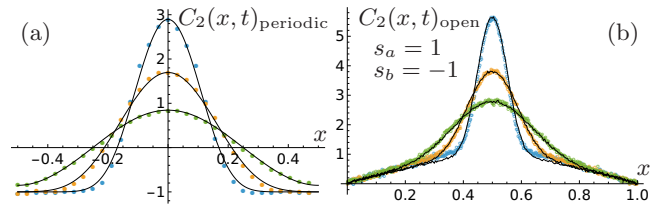
\begin{figure}
	\includegraphics[width=42mm]{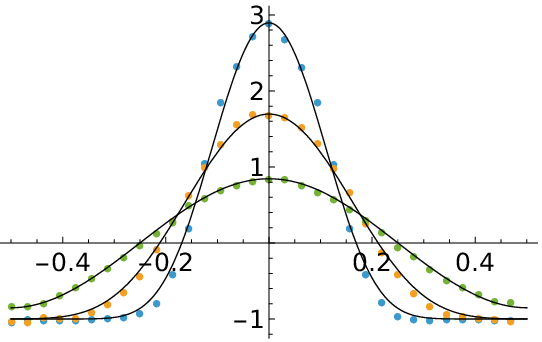}
	\includegraphics[width=42mm]{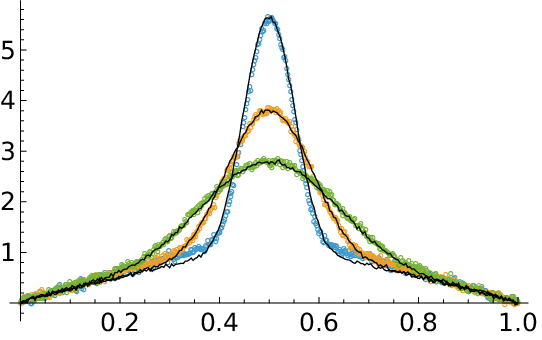}
	\begin{picture}(0,0)
		\put(-85,22){(a)}
		\put(-46,8.5){$x$}
		\put(-64,24){$C_{2}(x,t)_{\rm periodic}$}
		\put(-7,22){(b)}
		\put(-3,4){$x$}
		\put(-40,24){$C_{2}(x,t)_{\rm open}$}
		\put(-38,20){$s_{a}=1$}
		\put(-38,17){$s_{b}=-1$}
	\end{picture}
	\caption{Stationary two-point correlation function $C_{2}(x,t)$ plotted as a function of the position $x$ for a few values of time $t$, for periodic (a) and open (b) spin chains. The solid line is the KPZ prediction, for times $t\approx0.020, 0.037, 0.074$ in (a) and times $t\approx0.0056, 0.0118, 0.0228$ in (b), in order of decreasing amplitude. The dots correspond to numerical simulations of the KPLL model with $L=32$ spins, $T=30, 50, 90$ time steps and averaging over $10^{5}$ samples in (a), and with $L=511$ spins, $T=1000, 2000, 4000$ time steps (in addition to $2\times10^{5}$ preliminary time steps used to reach the stationary state) and averaging over $10^{7}$ samples in (b). In (b), the boundary magnetizations $s_{a}=1$, $s_{b}=-1$ defined in (\ref{Sab[sab]}) correspond to KPZ boundary densities $\sigma_{a}=-\sigma_{b}\approx2.4$.}
	\label{fig c2x}
\end{figure}

Alternatively, the presence of large finite size corrections at the boundaries, associated with the square root singularities in (\ref{C1 small}) below, makes it preferable in some cases to fix  $\sigma_{a}$, $\sigma_{b}$ from e.g. the total magnetization in both halves of the chain. The former scheme (\ref{eq boundaries}) was used for the magnetization profile in figure~\ref{fig C1}, the latter one for the two-point correlations of figure~\ref{fig c2x}. Additionally, we emphasize that for fixed boundary magnetizations $s_{a},s_{b}$, the averaged local magnetization $\langle m(x,t)\rangle$ in the spin chain depends on the parameter $\tau$ of the KPLL model, so that correspondence between $s_{a},s_{b}$ and the KPZ boundary densities $\sigma_{a},\sigma_{b}$ also depends on $\tau$. In particular, as explained below (\ref{local map KPLL}), we used $\tau=1$ in figure~\ref{fig C1} and $\tau=0.1$ in figure~\ref{fig c2x}.

The hypothesis (\ref{m[sigma_a,sigma_b]}) with two independent KPZ fields is tested below from numerical simulations of the KPLL and Ishimori spin chains. A good agreement is found both for the average magnetization in the stationary state and for the stationary two-point correlation function. Correlation functions of the magnetization $m(x,t)$ are compared with stationary KPZ correlation functions 
\begin{equation}
	\label{c1 c2}
	c_{1}(x)=\langle\sigma(x,t)\rangle_{\rm st}
	\;\,\text{and}\;\,
	c_{2}(x,t)=\langle\sigma(0,0)\,\sigma(x,t)\rangle_{\rm st}\,,
\end{equation}
where $_{\rm st}$ denotes stationary initial condition for the density field $\sigma(x,t)$ (but the height $h(x,t)$ grows forever with time, toward negative values by convention here). For periodic boundaries, one has $c_{1}(x)=0$, while the non-trivial function $c_{2}(x,t)$ has an exact Fourier representation \cite{P2016.1}. For open boundaries, an exact integral expression for $c_{1}(x)$ can be extracted from the known representation \cite{BLD2022.1,BWW2023.1} of the stationary state in terms of Brownian motions, see SM \cite{SM}. On the other hand, $c_{2}(x,t)$ is still unknown for open boundaries, but may be evaluated from numerical simulations of a discrete model in KPZ universality, see SM \cite{SM}.\\

\noindent\textit{Magnetization profile in the stationary state}. -- For the spin chain with periodic boundaries, one has $\langle S_{i}^{\rm z}\rangle_{\rm st}=0$. This is not generally the case for open boundaries where we define the stationary magnetization profile from (\ref{m[S]}) as
\begin{equation}
	\label{C_{1}[S]}
	C_{1}(x\,;s_{a},s_{b})=\langle m(x,t)\rangle_{\rm st}^{s_{a},s_{b}}\;,
\end{equation}
in the limit of a large number $L$ of spins and a long time $t\gg1$. The function $C_{1}$ is non-zero when $(s_{a},s_{b})\neq(0,0)$, and is antisymmetric under flipping both boundary magnetizations simultaneously, i.e. $C_{1}(x\,;-s_{a},-s_{b})=-C_{1}(x\,;s_{a},s_{b})$. According to the hypothesis (\ref{m[sigma_a,sigma_b]}), $C_{1}$ should be related to the average KPZ density field in the stationary state $c_{1}$ defined in (\ref{c1 c2}) as
\begin{equation}
	\label{C1[c1-c1] open}
	C_{1}(x\,;s_{a},s_{b})\simeq\frac{c_{1}(x\,;\sigma_{a},\sigma_{b})-c_{1}(x\,;-\sigma_{a},-\sigma_{b})}{\mathcal{N}(\sigma_{a},\sigma_{b})}\;.
\end{equation}
The boundary densities $\sigma_{a}$, $\sigma_{b}$ are computed in terms of the local magnetization at sites $i=1$, $i=L$ from (\ref{eq boundaries}). By symmetry, $s_{a}=s_{b}=0$ corresponds to $\sigma_{a}=\sigma_{b}=0$, where $C_{1}(x)=0$ for all $x$, and more generally $s_{a}+s_{b}=0$ corresponds to $\sigma_{a}+\sigma_{b}=0$, where $C_{1}(1-x)=-C_{1}(x)$.

From numerical simulations of the KPLL model, we find that the two sides of (\ref{C1[c1-c1] open}) agree quite well, although whether (\ref{C1[c1-c1] open}) becomes exact when $L\to\infty$ is not entirely clear from numerics, see figure~\ref{fig C1} for a representative case and SM \cite{SM} for a few other cases and for the exact KPZ expressions used to evaluate $c_{1}(x\,;\sigma_{a},\sigma_{b})$. We emphasize that a combination of two KPZ modes is needed~: taking a single KPZ mode with either sign for the asymmetry leads to a completely wrong prediction. As noted already in \cite{RDSK2024.1}, the magnetization profile does not simply interpolate linearly between $\sigma_{a}$ and $\sigma_{b}$, even when these parameters are small, where we find
\begin{equation}
	\label{C1 small}
	C_{1}(x)\simeq\frac{\sigma_{a}+\sigma_{b}}{2}+\frac{\sigma_{b}-\sigma_{a}}{\pi}\,\arcsin(2x-1)\;.
\end{equation}
We note that (\ref{C1 small}) essentially matches with the magnetization profile obtained in \cite{ARRS1999.1} within the light-cone for a bipartite quench of an infinitely long chain of free Fermions.

For large $\sigma_{a}$, $\sigma_{b}$, corresponding to large $s_{a}$, $s_{b}$, (\ref{m[sigma_a,sigma_b]}) predicts a non-trivial phase diagram \cite{SD1993.1,DEHP1993.1,BLD2022.1,BWW2023.1} with constant $C_{1}(x)$ proportional to $\sigma_{a}$, $\sigma_{b}$ or $\sigma_{a}+\sigma_{b}$ depending on their respective signs, see SM \cite{SM}, plus fluctuations such that the variance of $\int_{0}^{x}\dd y\,m(y,t)$ stays bounded. In the special case $s_{a}+s_{b}=0$, however, phase separation \cite{BD2005.1,BDSGJLL2005.1,KS2017.1,BKL2018.1,SNC2018.1} occurs, with $C_{1}(x)$ proportional to $\sigma_{a}$ ($\sigma_{b}$) when $0\leq x<u$ ($u<x\leq1$) and $u$ distributed uniformly between $0$ and $1$. Then, for large $s_{a}=-s_{b}$, the variance of $\int_{0}^{x}\dd y\,m(y,t)$ and hence $\sqrt{L}\,\Var(M)_{\rm st}$ is instead expected to be divergent, which we indeed observe in our numerical simulations of the KPLL model, see figure~\ref{fig C1}c. The complete phase diagram at large $s_{a}$, $s_{b}$ is more difficult to check since a large number of spins is needed in order to preserve the scaling relation (\ref{Sab[sab]}).\\

\noindent\textit{Stationary two-point correlation function}. -- Even though we consider a stationary initial condition for the spin chain, non-trivial correlations in time exist, which are captured by
\begin{equation}
	\label{C_{2}[S]}
	C_{2}(x,t)=\langle m(x_{0},0)\,m(x_{0}+x,t)\rangle_{\rm st}\;,
\end{equation}
where $m(x,t)$ is the local magnetization defined in (\ref{m[S]}). For the periodic chain, $C_{2}(x,t)$ is independent of $x_{0}$, and we additionally average over it in our simulations. For the open chain, we consider in the following only correlations from the middle of the chain, with $x_{0}=1/2$.

Unlike for the stationary magnetization profile, which is independent of time, an important issue is now how to fix the KPZ time scale $t$ from the time $T$ of the spin chain. In principle, one needs for each value of $L$ a single data point at a chosen time $T$ in order to fix $t$ from the scaling relation $T\sim tL^{3/2}$. In practice, however, the collapse as a function of $T/L^{3/2}$ for large $L$ is quite slow, and requires an increasingly large number $L$ of spins for higher values of $t$. It is enough for our purpose to fix the KPZ time \emph{at each time step $T$} by matching the value of $C_{2}(0,t)$ with the KPZ prediction. Then, the remaining spatial dependency for the spin chain can be compared with the KPZ prediction. We emphasize that apart from $t$ (and the boundary densities $\sigma_{a}$, $\sigma_{b}$ set from stationary fluctuations for the open chain), \emph{no extra adjustable parameter} is available, so that the family of curves $C_{2}(x,t)$ indexed by $t$ is defined unambiguously.

We begin with the spin chain with periodic boundaries, where we expect that $C_{2}(x,t)\simeq c_{2}(x,t)$ with $c_{2}$ the stationary two-point function of a single KPZ field with periodic boundaries defined in (\ref{c1 c2}). The function $c_{2}(x,t)$ describes the crossover between KPZ on the infinite line, recovered at short time when boundary effects can be neglected since the correlation length is still small compared to the system size, and a non-equilibrium stationary state with correlations spanning the whole system at late time. In particular, when $t\to0$, one has
\begin{equation}
	\label{c2PS}
	t^{2/3}c_{2}(t^{2/3}y,t)\to c_{2}^{\rm line}(y)
\end{equation}
with $c_{2}^{\rm line}$ the Pr\"ahofer-Spohn scaling function \cite{PS2004.1+} characterizing KPZ fluctuations on the infinite line.

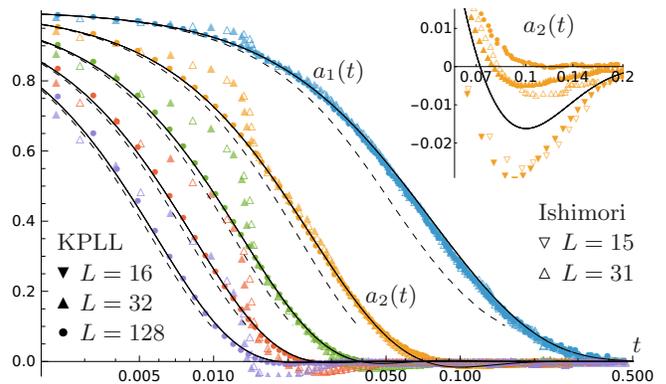
\begin{figure}
	\includegraphics[width=86mm]{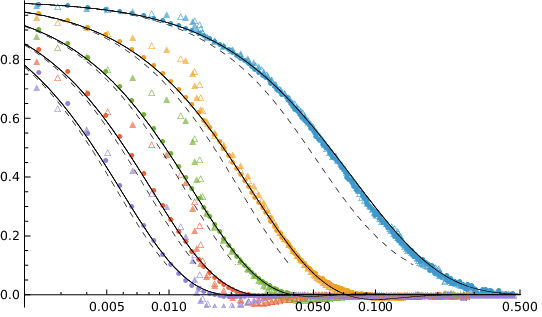}
	\begin{picture}(0,0)
		\put(10,31.5){\includegraphics[width=30mm]{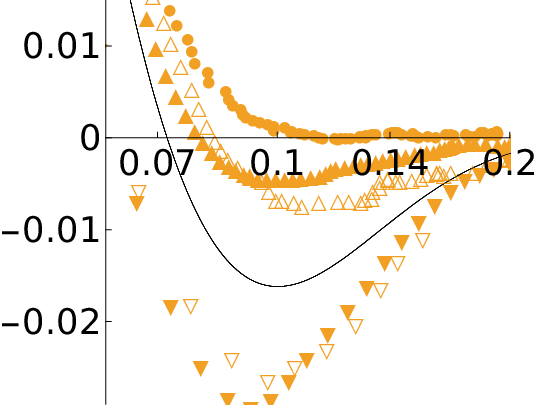}}
		\put(39.5,8.5){$t$}
		\put(-3,45){$a_{1}(t)$}
		\put(4,15){$a_{2}(t)$}
		\put(25,51){$a_{2}(t)$}
		\put(-37,10){
			\put(0,12){KPLL}
			\put(0,7){\includegraphics[width=2mm]{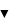}}\put(3,8){$L=16$}
			\put(0,3){\includegraphics[width=2mm]{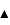}}\put(3,4){$L=32$}
			\put(0,-1){\includegraphics[width=2mm]{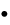}}\put(3,0){$L=128$}
		}
		\put(27,18){
			\put(0,8){Ishimori}
			\put(0,3){\includegraphics[width=2mm]{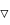}}\put(3,4){$L=15$}
			\put(0,-1){\includegraphics[width=2mm]{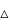}}\put(3,0){$L=31$}
		}
	\end{picture}
	\caption{First Fourier coefficients $a_{1}(t),\ldots,a_{5}(t)$ (from top to bottom) of the stationary two-point function $C_{2}(x,t)\simeq c_{2}(x,t)$ with periodic boundaries, plotted as a function of time. The solid lines are the exact results for the $a_{p}(t)$, and the dashed lines the short time approximation $\hat{c}_{2}^{\rm line}(2\pi k/\ell)$. The symbols represent the data from numerical simulations of the KPLL model (Ishimori chain) averaged over $10^{6}$ ($10^{5}$) samples. Inset: zoom around the first zero of $a_{2}(t)$.}
	\label{fig KPLL Fourier P 32 128 lowS}
\end{figure}

As a function of the position $x$ for fixed times $t$, we observe a good agreement between the KPZ exact result for $c_{2}(x,t)$ with periodic boundaries stated in SM \cite{SM} and the numerical simulations for the KPLL model, already for $L=32$ spins see figure~\ref{fig c2x}. As a function of time, we consider rather the Fourier coefficients in space of $c_{2}$. Density conservation implies that the zeroth coefficient vanishes, and we write
\begin{equation}
	\label{c2xt[at] P}
	c_{2}(x,t)=2\sum_{k=1}^{\infty}a_{k}(t)\cos(2\pi kx)\;.
\end{equation}
One has $c_{2}(x,0)=-1$ for $x\neq0$, corresponding to $a_{k}(0)=1$ for all $k$. Additionally (\ref{c2PS}) implies that $c_{2}(x,t)$ should be well approximated for $-1/2<x<1/2$ by $\ell\,c_{2}^{\rm line}(\ell x)$ with $\ell=t^{-2/3}$, which leads to $a_{k}(t)\simeq\hat{c}_{2}^{\rm line}(2\pi k/\ell)$ with $\hat{c}_{2}^{\rm line}$ the Fourier transform of $c_{2}^{\rm line}$. Long enough times are then necessary to observe the emergence of finite size effects, see figure~\ref{fig KPLL Fourier P 32 128 lowS}. On the other hand, the $a_{k}(t)$ converge exponentially fast to $0$ at late times, with relaxation time $t_{k}^{*}$ decreasing with $k$, see SM \cite{SM}. In particular, from times $t\approx0.05$, one has $c_{2}(x,t)\approx a_{1}(t)\cos(2\pi x)$.

Comparing the first Fourier coefficients for the KPLL and Ishimori spin chains with the exact result for KPZ, we find overall an excellent agreement. Quite remarkably, this holds even for relatively small chains with a few dozen spins, where we observe however a noticeable ``bump'' compared to the KPZ prediction, see figure~\ref{fig KPLL Fourier P 32 128 lowS}, caused by the marked microscopic spatial structure at shorter times which prevents fixing $t$ from $C_{2}(0,t)$ in a reliable way, see SM \cite{SM}. For larger values of $L$, the bump is pushed to shorter time $t$ as the range of validity of KPZ universality extends to smaller values of $t$, and its amplitude becomes smaller, see SM \cite{SM}.

On closer inspection, see inset of figure~\ref{fig KPLL Fourier P 32 128 lowS}, we observe a small deviation from the KPZ exact result for $a_{2}(t)$ around the time $t\approx0.1$ where it changes sign and reaches its largest negative value. This deviation appears to persist at least for moderate number $L$ of spins. Since it is located around a point where $a_{2}(t)$ vanishes, a possible explanation might be that subleading terms in $L$ simply become dominant in this region for moderate values of $L$, but do eventually converge to zero for large enough $L$. Another possibility would be that this deviation signals either the existence of a non-zero coupling between the two KPZ fields or even the presence of additional fields.

We consider now the case of open boundaries, where the reservoirs at both ends impose their magnetization to the spin chain, and restrict for simplicity to $s_{a}+s_{b}=0$. We compare correlations $C_{2}(x,t)$ from the middle of the chain with the exact KPZ result, accounting for two KPZ fields as in (\ref{m[sigma_a,sigma_b]}). Writing $c_{2}(x,t;\sigma_{a},\sigma_{b})=\langle\sigma^{\sigma_{a},\sigma_{b}}(1/2,0)\,\sigma^{\sigma_{a},\sigma_{b}}(x,t)\rangle$, we expect $C_{2}(x,t)\simeq(c_{2}(x,t;\sigma_{a},\sigma_{b})+c_{2}(x,t;-\sigma_{a},-\sigma_{b}))/\mathcal{N}(\sigma_{a},\sigma_{b})^{2}$ since $C_{1}(1/2)=0$ when $s_{a}+s_{b}=0$.

We use the same procedure as for periodic boundaries in order to compare simulations of the KPLL model with the KPZ prediction. Since no exact result is available in that case, we rely on numerical simulations of a known model of classical hard-core particles in KPZ universality in order to extract the universal scaling function, see SM \cite{SM}.

We observe again a good agreement for the stationary two point function between the spin chain and the KPZ prediction, although the convergence for large number $L$ of spins appears to be slower than for periodic boundaries, and at least a few hundred spins appear to be needed in order to reach reasonably good convergence. The family of functions $x\mapsto C_{2}(x,t)$ indexed by time $t$ obtained for open boundaries have a markedly different shape than for periodic boundaries, see figure~\ref{fig c2x}.\\

\noindent\textit{Conclusions}. -- By considering the time scale where fluctuations relax to a stationary state, we have obtained substantially more evidence for the presence of KPZ fluctuations in isotropic classical integrable spin chains. Considering carefully boundary effects on these time scales, we observe that relatively few spins are actually required to observe universal KPZ scaling functions with good accuracy, especially with periodic boundaries. Most importantly, the observation of stationary KPZ fluctuations for open chains with boundary magnetizations scaled down to typical fluctuations within the chain opens a clear path to analytic computations, which should help to clarify the nature of fluctuations in integrable spin chains. In particular whether the KPZ picture put forward in this letter for spin chains with boundaries becomes exact in the limit of a large number of spins, or whether the small deviations we observed persist.\\

Acknowledgements: it is a pleasure to thank T.~Prosen and H.~Spohn for inspiring discussions on classical spin chains. Hospitality at CMSA Harvard, where this study was initiated, is gratefully acknowledged. This study was partially supported through the grant NanoX n° ANR-17-EURE-0009 in the framework of the ``Programme des Investissements d’Avenir'', and through the F\'ed\'eration de recherche Mati\`ere et Interactions at the University of Toulouse.


\end{document}


\title{Supplemental material to the paper \\ ``Cheaper access to universal fluctuations in integrable spin chains\\from boundary effects''}
\author{Sylvain Prolhac}
\email{sylvain.prolhac@irsamc.ups-tlse.fr}
\affiliation{Laboratoire de Physique Th\'eorique, Universit\'e de Toulouse, France}
\maketitle
\tableofcontents

\clearpage
\section{Stationary state for KPZ}

\subsection{Brownian representation}
The stationary KPZ height $h_{\rm st}(x)=\int_{0}^{x}\dd y\,\sigma_{\rm st}(y)$ is a continuous random function of $x$, $0\leq x\leq 1$, with $h_{\rm st}(0)=0$. It is typically non-differentiable everywhere. With periodic boundaries, $h_{\rm st}$ is simply a standard Brownian bridge, i.e. $h_{\rm st}(x)=w(x)-xw(1)$ with $w$ a standard Brownian motion.

For KPZ with open boundaries and arbitrary boundary densities $\sigma_{a}$ and $\sigma_{b}$, one has the marginal representation $h_{\rm st}^{\sigma_{a},\sigma_{b}}(x)=w(x)$, where two independent standard Brownian motions $w$ and $\tilde{w}$ are re-weighted by
\begin{equation}
	\ee^{(\sigma_{a}-\sigma_{b})\max(w-\tilde{w})}\,\ee^{\sigma_{b}(w(1)-\tilde{w}(1))}\;.
\end{equation}
An explicit integral representation for $\langle h_{\rm st}^{\sigma_{a},\sigma_{b}}(x)\rangle$ follows, see below.

\subsection{Exact formulas for some stationary expectation values for KPZ with open boundaries}
Moments of the stationary KPZ height field with boundary densities $\sigma_{a}$ and $\sigma_{b}$ (the notations $u=\sigma_{a}$ and $v=-\sigma_{b}$ are used instead in the mathematical literature) can be expressed in terms of Brownian averages as
\begin{equation}
	\langle h_{\rm st}(x)^{k}\rangle=\frac{Z_{k}(x)}{Z_{0}}
\end{equation}
with
\begin{equation}
	Z_{k}(x)=\langle w(x)^{k}\,\ee^{(\sigma_{a}-\sigma_{b})\max(w-\tilde{w})}\,\ee^{\sigma_{b}(w(1)-\tilde{w}(1))}\rangle
\end{equation}
and $Z_{0}=Z_{0}(x)$ independent of $x$. The independent standard Brownian motions (Wiener processes) $w(u)$ and $\tilde{w}(u)$, $u\in[0,1]$, verify $w(0)=\tilde{w}(0)=0$. The maximum of $w(u)-\tilde{w}(u)$ is computed in the interval $u\in[0,1]$.

\subsubsection{Explicit formula for $Z_{0}$}
Since $\frac{w-\tilde{w}}{\sqrt{2}}$ is also a Wiener process, the Brownian expectation value for $Z_{0}$ can be computed as
\begin{equation}
	Z_{0}=\int_{0}^{\infty}\dd z\int_{-\infty}^{z}\dd y\,\ee^{\sqrt{2}(\sigma_{a}-\sigma_{b})z+\sqrt{2}\sigma_{b}y}\partial_{z}p_{1,z}(y)
\end{equation}
where the probability density
\begin{equation}
	p_{x,z}(y)=\frac{\P(w(x)\in[y,y+\dd y],\max(w)<z)}{\dd y}
\end{equation}
has the explicit expression
\begin{equation}
	p_{x,z}(y)=\frac{\ee^{-\frac{y^{2}}{2x}}}{\sqrt{2\pi x}}-\frac{\ee^{-\frac{(2z-y)^{2}}{2x}}}{\sqrt{2\pi x}}
\end{equation}
from the method of images. One finds
\begin{equation}
	Z_{0}=\frac{\sigma_{a}\,\ee^{\sigma_{a}^{2}}\erfc(-\sigma_{a})+\sigma_{b}\,\ee^{\sigma_{b}^{2}}\erfc(\sigma_{b})}{\sigma_{a}+\sigma_{b}}\;,
\end{equation}
where $\erfc(u)=\frac{2}{\sqrt{\pi}}\int_{u}^{\infty}\dd v\,\ee^{-v^{2}}$ is the complementary error function. This reduces to $Z_{0}=\ee^{\sigma_{a}^{2}}$ when $\sigma_{a}=\sigma_{b}$, and to $Z_{0}=\frac{2\sigma_{a}}{\sqrt{\pi}}+(1+2\sigma_{a}^{2})\,\ee^{\sigma_{a}^{2}}\erfc(-\sigma_{a})$ when $\sigma_{a}+\sigma_{b}=0$.

\subsubsection{Integral formula for $Z_{1}(x)$}
In order to compute $Z_{k}(x)$, we factorize the joint probability of $w$ and $\tilde{w}$ at $x$ as
\begin{align}
	&\frac{\P((\Delta w)_{0}^{x}<z_{1},w(x)\in\dd y_{1},\tilde{w}(x)\in\dd\tilde{y}_{1},(\Delta w)_{x}^{1}<z_{2},w(1)\in\dd y_{2},\tilde{w}(1)\in\dd\tilde{y}_{2})}{\dd y_{1}\dd\tilde{y}_{1}\dd y_{2}\dd\tilde{y}_{2}}\nonumber\\
	&=p_{x,z_{1}}(y_{1},\tilde{y}_{1})\,p_{1-x,z_{2}-y_{1}+\tilde{y}_{1}}(y_{2}-y_{1},\tilde{y}_{2}-\tilde{y}_{1})\;,
\end{align}
using the Markov property and the translation invariance of the Wiener process, with the shorthand $w\in\dd y$ for $w\in[y,y+\dd y]$, and $(\Delta w)_{a}^{b}<z$ for $\forall u\in[a,b],w(u)-\tilde{w}(u)<z$. The probability density
\begin{equation}
	p_{x,z}(y,\tilde{y})=\frac{\P(\forall u\in[0,x],w(u)-\tilde{w}(u)<z,w(x)\in[y,y+\dd y],\tilde{w}(x)\in[\tilde{y},\tilde{y}+\dd\tilde{y}])}{\dd y\,\dd\tilde{y}}
\end{equation}
with $z>0$ is the unique solution of the partial differential equation $\partial_{x}p_{x,z}(y,\tilde{y})=\frac{1}{2}(\partial_{y}^{2}+\partial_{\tilde{y}}^{2})p_{x,z}(y,\tilde{y})$ with initial condition $p_{0,z}(y,\tilde{y})=\delta(y)\delta(\tilde{y})$ and boundary condition $p_{x,z}(y,\tilde{y})=0$ when $y-\tilde{y}=z$. The method of images gives the explicit expression
\begin{equation}
	p_{x,z}(y,\tilde{y})=\frac{\ee^{-\frac{y^{2}+\tilde{y}^{2}}{2x}}}{\sqrt{2\pi x}}-\frac{\ee^{-\frac{(y-z)^{2}+(\tilde{y}+z)^{2}}{2x}}}{\sqrt{2\pi x}}\;.
\end{equation}
Then,
\begin{align}
	Z_{k}(x)=\int_{0}^{\infty}\dd z_{1}\int_{-\infty}^{\infty}\dd z_{2}\,\dd y_{1}\,\dd\tilde{y}_{1}\,\dd y_{2}\,\dd\tilde{y}_{2}\,&1_{\{y_{1}-\tilde{y}_{1}<z_{1}\}}\,1_{\{y_{1}-\tilde{y}_{1}<z_{2}\}}\,1_{\{y_{2}-\tilde{y}_{2}<z_{2}\}}\,y_{1}^{k}\,\ee^{(\sigma_{a}-\sigma_{b})\max(z_{1},z_{2})}\,\ee^{\sigma_{b}(y_{2}-\tilde{y}_{2})}\nonumber\\
	&\times\partial_{z_{1}}p_{z_{1}}(x,y_{1},\tilde{y}_{1})\,\partial_{z_{2}}p_{z_{2}-y_{1}+\tilde{y}_{1}}(1-x,y_{2}-y_{1},\tilde{y}_{2}-\tilde{y}_{1})\;.
\end{align}
Computing some integrals, we eventually obtain
\begin{align}
	Z_{1}(x)=\int_{0}^{\infty}\dd y\,\Bigg(
	-\frac{A_x'(y) B_{1-x}'(y)}{(\sigma_{a}+\sigma_{b})^{2}}
	&+\frac{(2 x \sigma_{a}^{2}-2 y \sigma_{a}+1) A_{1-x}'(y) A_x'(y)}{2 \sigma_{a} (\sigma_{a}+\sigma_{b})}
	+\frac{(-2 x \sigma_{b}^{2} (\sigma_{a}+\sigma_{b})-\sigma_{a}+\sigma_{b}) B_{1-x}'(y) B_x'(y)}{2 \sigma_{b} (\sigma_{a}+\sigma_{b})^{2}}\nonumber\\
	&-\frac{\ee^{-\frac{y^{2}}{4 x}} (y \sigma_{a}-1) A_{1-x}'(y)}{2 \sqrt{\pi x} \sigma_{a} (\sigma_{a}+\sigma_{b})}
	-\frac{\ee^{-\frac{y^{2}}{4 x}} (y \sigma_{b}+1) B_{1-x}'(y)}{2 \sqrt{\pi x} \sigma_{b} (\sigma_{a}+\sigma_{b})}
	\Bigg)&
\end{align}
with $^\prime$ denoting derivation with respect to $y$ and
\begin{align}
	A_{u}(y)&=\ee^{u\,\sigma_{a}^{2}-y\,\sigma_{a}}\erfc\Big(\frac{y-2u\,\sigma_{a}}{2\sqrt{u}}\Big)\\
	B_{u}(y)&=\ee^{u\,\sigma_{b}^{2}+y\,\sigma_{b}}\erfc\Big(\frac{y+2u\,\sigma_{b}}{2\sqrt{u}}\Big)\;.
\end{align}

\begin{figure}
	\hspace*{-1mm}
	\begin{tabular}{lll}
		\begin{tabular}{l}
			\includegraphics[width=42mm]{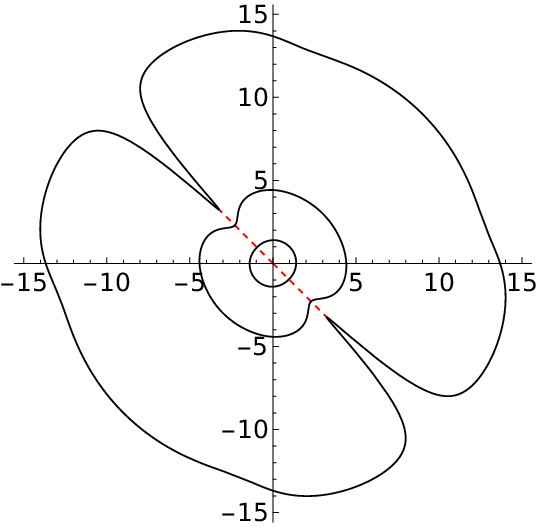}
			\begin{picture}(0,0)
				\put(-43,37){(a)}
				\put(-5.5,22){$m_{a}$}
				\put(-21,39.2){$m_{b}$}
			\end{picture}
		\end{tabular}
		&\hspace*{-5mm}&
		\begin{tabular}{l}
			\includegraphics[width=42mm]{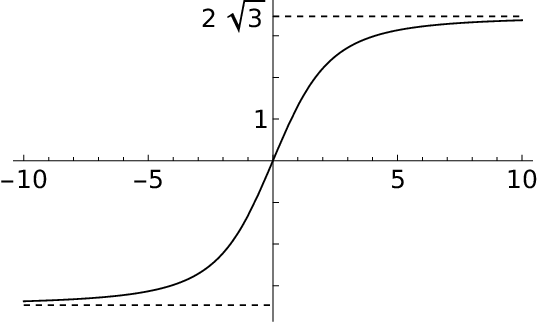}
			\begin{picture}(0,0)
				\put(-5,29){(b)}
				\put(-4,14){$\sigma_{a}$}
				\put(-21,25){$m_{a}$}
				\put(-30,-5){$\sigma_{a}+\sigma_{b}=0$}
			\end{picture}
		\end{tabular}
	\end{tabular}
	\caption{Plot of $m_{a}$ and $m_{b}$ defined in (\ref{ma[sigma_ab]}), (\ref{mb[sigma_ab]}), from which KPZ boundary densities $\sigma_{a}$ and $\sigma_{b}$ are computed for the spin chain. In (a), the solid black curves (dashed red line) are the locus of $(m_{a},m_{b})$ for $(\sigma_{a},\sigma_{b})$ on the circles $(\sigma_{a}^{2}+\sigma_{b}^{2})^{1/2}=1,3,9$ (on the line $\sigma_{a}+\sigma_{b}=0$). In (b), $m_{a}$ is plotted as a function of $\sigma_{a}$ on the line $\sigma_{a}+\sigma_{b}=0$.}
	\label{fig walnut}
\end{figure}
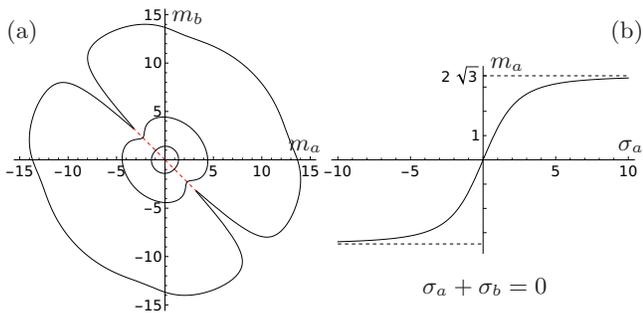

\subsubsection{Explicit formulas at $x=1$}
Calculations simplify at $x=1$, where
\begin{equation}
	Z_{k}(1)=\int_{0}^{\infty}\dd z\int_{-\infty}^{\infty}\dd y\,\dd\tilde{y}\,1_{\{y-\tilde{y}<z\}}\,y^{k}\,\ee^{(\sigma_{a}-\sigma_{b})z}\,\ee^{\sigma_{b}(y-\tilde{y})}\partial_{z}p_{z}(1,y,\tilde{y})\;.
\end{equation}
One finds in particular the explicit expression for the mean
\begin{equation}
\langle h_{\rm st}^{\sigma_{a},\sigma_{b}}(1)\rangle=\frac{\frac{\sigma_{a}-\sigma_{b}}{\sqrt{\pi }}+\big(\sigma_{a}^{2}-\frac{\sigma_{a}-\sigma_{b}}{2(\sigma_{a}+\sigma_{b})}\big)\,\ee^{\sigma_{a}^{2}}\erfc(-\sigma_{a})+\big(\sigma_{b}^{2}+\frac{\sigma_{a}-\sigma_{b}}{2(\sigma_{a}+\sigma_{b})}\big)\,\ee^{\sigma_{b}^{2}}\erfc(\sigma_{b})}{\sigma_{a}\,\ee^{\sigma_{a}^{2}}\erfc(-\sigma_{a})+\sigma_{b}\,\ee^{\sigma_{b}^{2}}\erfc(\sigma_{b})}\;,
\end{equation}
which implies $\langle h_{\rm st}^{\sigma,\sigma}(1)\rangle=\sigma$ and $\langle h_{\rm st}^{\sigma,-\sigma}(1)\rangle=0$, and for the variance
\begin{align}
	\Var(h_{\rm st}^{\sigma_{a},\sigma_{b}}(1))=1+\frac{1}{(\sigma_{a}+\sigma_{b})^{2}}&+\frac{1}{(\sigma_{a}\,\ee^{\sigma_{a}^{2}}\erfc(-\sigma_{a})+\sigma_{b}\,\ee^{\sigma_{b}^{2}}\erfc(\sigma_{b}))^{2}}\Bigg(\frac{1-\sigma_{a}\sigma_{b}-(\sigma_{a}-\sigma_{b})^{2}}{\pi}\\
	&-\frac{(\ee^{\sigma_{a}^{2}} \erfc(-\sigma_{a})+\ee^{\sigma_{b}^{2}}\erfc(\sigma_{b}))^{2}}{4}
	-\frac{(\sigma_{a}-\sigma_{b})(\sigma_{a}^{2}\,\ee^{\sigma_{a}^{2}}\erfc(-\sigma_{a})+\sigma_{b}^{2}\,\ee^{\sigma_{b}^{2}}\erfc(\sigma_{b}))}{\sqrt{\pi }}\nonumber\\
	&+\pi(\sigma_{a} \sigma_{b}-1)\Big(\frac{1}{\pi}+\frac{(\sigma_{a}-\sigma_{b})\,\ee^{\sigma_{a}^{2}}\erfc(-\sigma_{a})}{\sqrt{\pi }}\Big)\Big(\frac{1}{\pi }+\frac{(\sigma_{a}-\sigma_{b})\,\ee^{\sigma_{b}^{2}}\erfc(\sigma_{b})}{\sqrt{\pi}}\Big)\Bigg)\;,\nonumber
\end{align}
which implies $\Var(h_{\rm st}^{\sigma,\sigma}(1))=1$ and
\begin{equation}
	\Var(h_{\rm st}^{\sigma,-\sigma}(1))=\frac{2\sigma(\sigma^{2}+4)+\sqrt{\pi}(3+9\sigma^{2}+2\sigma^{4})\,\ee^{\sigma^{2}}\erfc(-\sigma)}{6\sigma+3\sqrt{\pi}(1+2\sigma^{2})\,\ee^{\sigma^{2}}\erfc(-\sigma)}\;.
\end{equation}
The functions
\begin{align}
	\label{ma[sigma_ab]}
	m_{a}(\sigma_{a},\sigma_{b})&=\frac{2\sigma_{a}}{\sqrt{\Var(h_{\rm st}^{\sigma_{a},\sigma_{b}}(1))+\Var(h_{\rm st}^{-\sigma_{a},-\sigma_{b}}(1))}}\\
	\label{mb[sigma_ab]}
	m_{b}(\sigma_{a},\sigma_{b})&=\frac{2\sigma_{b}}{\sqrt{\Var(h_{\rm st}^{\sigma_{a},\sigma_{b}}(1))+\Var(h_{\rm st}^{-\sigma_{a},-\sigma_{b}}(1))}}
\end{align}
are used in the letter to fix the boundary densities $\sigma_{a}$ and $\sigma_{b}$ from the boundary magnetizations (normalized by the variance of the total magnetization).

\subsubsection{Expansion for small $\sigma_{a},\sigma_{b}$}
For small values of $\sigma_{a}$ and $\sigma_{b}$, one finds
\begin{equation}
	\Var(h_{\rm st}^{\sigma_{a},\sigma_{b}}(1))\simeq1+\frac{\sigma_{a}-\sigma_{b}}{3\sqrt{\pi}}+\frac{3\pi-8}{12\pi}\,(\sigma_{a}-\sigma_{b})^{2}\;,
\end{equation}
which leads to the map
\begin{align}
	m_{a}(\sigma_{a},\sigma_{b})&\simeq\sqrt{2}\,\sigma_{a}-\frac{3\pi-8}{12\sqrt{2}\,\pi}\,\sigma_{a}\,(\sigma_{a}-\sigma_{b})^{2}\\
	m_{b}(\sigma_{a},\sigma_{b})&\simeq\sqrt{2}\,\sigma_{b}-\frac{3\pi-8}{12\sqrt{2}\,\pi}\,\sigma_{b}\,(\sigma_{a}-\sigma_{b})^{2}
\end{align}
between boundary densities and normalized boundary magnetizations. The height profile behaves as
\begin{equation}
	\langle h_{\rm st}^{\sigma_{a},\sigma_{b}}(x)\rangle\simeq\sigma_{a}x+\frac{\sigma_{b}-\sigma_{a}}{\pi}\Big(\sqrt{x(1-x)}+(2x-1)\arcsin\sqrt{x}\Big)\;,
\end{equation}
which leads to the density profile
\begin{equation}
	\label{magnetization profile small}
	\langle\sigma_{\rm st}^{\sigma_{a},\sigma_{b}}(x)\rangle
	=\partial_{x}\langle h_{\rm st}^{\sigma_{a},\sigma_{b}}(x)\rangle
	\simeq\frac{\sigma_{a}+\sigma_{b}}{2}+\frac{\sigma_{b}-\sigma_{a}}{\pi}\,\arcsin(2x-1)\;.
\end{equation}

\subsubsection{Large $\sigma_{a},\sigma_{b}$ asymptotics}
\begin{figure}
	\begin{center}
		\begin{tabular}{lll}
			\begin{tabular}{l}
				\includegraphics[width=60mm]{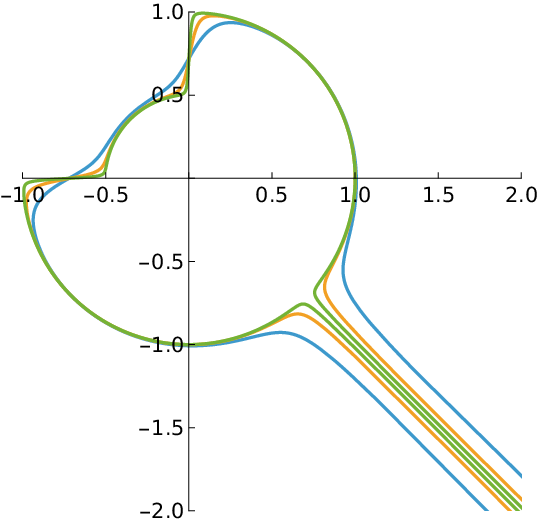}
				\begin{picture}(0,0)
					\put(-5,41){$\sigma_{a}$}
					\put(-39,59){$\sigma_{b}$}
				\end{picture}
			\end{tabular}
			&\hspace*{10mm}&
			\begin{tabular}{l}
				\begin{picture}(52,56)
					\put(14,55){$\Var(h_{\rm st}^{\sigma_{a},\sigma_{b}}(1))$}
					\put(0,0){\color[rgb]{0.9,0.9,1}\polygon*(0,25)(25,25)(25,50)(0,50)}
					\put(0,0){\color[rgb]{1,0.9,0.9}\polygon*(25,25)(25,50)(50,50)(50,0)}
					\put(0,0){\color[rgb]{0.9,1.0,0.9}\polygon*(25,25)(0,25)(0,0)(50,0)}
					\put(0,25){\vector(1,0){52}}\put(47,26){$\sigma_{a}$}
					\put(25,0){\vector(0,1){52}}\put(25.8,48){$\sigma_{b}$}
					\put(25,25){\thicklines\line(1,-1){25}}
					\put(5,42){$1/2$}
					\put(37,32){$1$}
					\put(15,11){$1$}
					\put(38,7){\rotatebox{-45}{$\sigma_{a}^{2}/3$}}
				\end{picture}
			\end{tabular}
		\end{tabular}
	\end{center}
	\caption{On the left, $z\Var(h_{\rm st}^{\sigma_{a},\sigma_{b}}(1))$ plotted in the complex plane with $z=\sigma_{a}+\ii\sigma_{b}$ and $|z|=10$ (blue) $|z|=30$ (orange) and $|z|=100$ (green). On the right, corresponding asymptotic behaviour for the variance of the KPZ height at $x=1$ when $\sigma_{a},\sigma_{b}$ are large.}
	\label{fig Var h}
\end{figure}

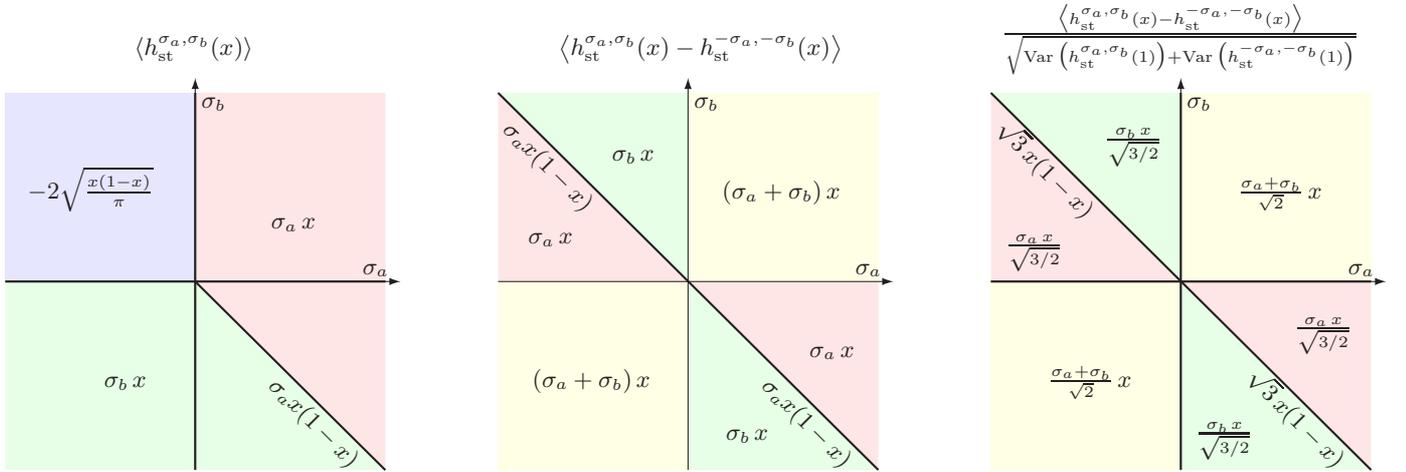
\begin{figure}
	\begin{center}
		\begin{tabular}{lllll}
			\begin{picture}(52,56)
				\put(17,55){$\langle h_{\rm st}^{\sigma_{a},\sigma_{b}}(x)\rangle$}
				\put(0,0){\color[rgb]{0.9,0.9,1}\polygon*(0,25)(25,25)(25,50)(0,50)}
				\put(0,0){\color[rgb]{1,0.9,0.9}\polygon*(25,25)(25,50)(50,50)(50,0)}
				\put(0,0){\color[rgb]{0.9,1.0,0.9}\polygon*(25,25)(0,25)(0,0)(50,0)}
				\put(0,25){\vector(1,0){52}}\put(47,26){$\sigma_{a}$}
				\put(25,0){\vector(0,1){52}}\put(25.8,48){$\sigma_{b}$}
				\put(0,25){\thicklines\line(1,0){50}}
				\put(25,0){\thicklines\line(0,1){50}}
				\put(25,25){\thicklines\line(1,-1){25}}
				\put(3,36){$-2\sqrt{\frac{x(1-x)}{\pi}}$}
				\put(35,32){$\sigma_{a}\,x$}
				\put(13,11){$\sigma_{b}\,x$}
				\put(34,11){\rotatebox{-45}{$\sigma_{a}x(1-x)$}}
			\end{picture}
			&\hspace*{10mm}&
			\begin{picture}(52,56)
				\put(8,55){$\big\langle h_{\rm st}^{\sigma_{a},\sigma_{b}}(x)-h_{\rm st}^{-\sigma_{a},-\sigma_{b}}(x)\big\rangle$}
				\put(0,0){\color[rgb]{1,0.9,0.9}\polygon*(25,25)(50,25)(50,0)}
				\put(0,0){\color[rgb]{1,0.9,0.9}\polygon*(25,25)(0,25)(0,50)}
				\put(0,0){\color[rgb]{0.9,1,0.9}\polygon*(25,25)(25,50)(0,50)}
				\put(0,0){\color[rgb]{0.9,1,0.9}\polygon*(25,25)(25,0)(50,0)}
				\put(0,0){\color[rgb]{1,1,0.9}\polygon*(25,25)(0,25)(0,0)(25,0)}
				\put(0,0){\color[rgb]{1,1,0.9}\polygon*(25,25)(50,25)(50,50)(25,50)}
				\put(0,25){\vector(1,0){52}}\put(47,26){$\sigma_{a}$}
				\put(25,0){\vector(0,1){52}}\put(25.8,48){$\sigma_{b}$}
				\put(0,50){\thicklines\line(1,-1){50}}
				\put(4,30){$\sigma_{a}\,x$}
				\put(41,15){$\sigma_{a}\,x$}
				\put(15,41){$\sigma_{b}\,x$}
				\put(30,4){$\sigma_{b}\,x$}
				\put(29.5,36){$(\sigma_{a}+\sigma_{b})\,x$}
				\put(4.5,11){$(\sigma_{a}+\sigma_{b})\,x$}
				\put(0,45){\rotatebox{-45}{$\sigma_{a}x(1-x)$}}
				\put(34,11){\rotatebox{-45}{$\sigma_{a}x(1-x)$}}
			\end{picture}
			&\hspace*{10mm}&
			\begin{picture}(52,56)
				\put(1.5,57){$\frac{\big\langle h_{\rm st}^{\sigma_{a},\sigma_{b}}(x)-h_{\rm st}^{-\sigma_{a},-\sigma_{b}}(x)\big\rangle}{\sqrt{\Var\big(h_{\rm st}^{\sigma_{a},\sigma_{b}}(1)\big)+\Var\big(h_{\rm st}^{-\sigma_{a},-\sigma_{b}}(1)\big)}}$}
				\put(0,0){\color[rgb]{1,0.9,0.9}\polygon*(25,25)(50,25)(50,0)}
				\put(0,0){\color[rgb]{1,0.9,0.9}\polygon*(25,25)(0,25)(0,50)}
				\put(0,0){\color[rgb]{0.9,1,0.9}\polygon*(25,25)(25,50)(0,50)}
				\put(0,0){\color[rgb]{0.9,1,0.9}\polygon*(25,25)(25,0)(50,0)}
				\put(0,0){\color[rgb]{1,1,0.9}\polygon*(25,25)(0,25)(0,0)(25,0)}
				\put(0,0){\color[rgb]{1,1,0.9}\polygon*(25,25)(50,25)(50,50)(25,50)}
				\put(0,25){\vector(1,0){52}}\put(47,26){$\sigma_{a}$}
				\put(25,0){\vector(0,1){52}}\put(25.8,48){$\sigma_{b}$}
				\put(0,25){\thicklines\line(1,0){50}}
				\put(25,0){\thicklines\line(0,1){50}}
				\put(0,50){\thicklines\line(1,-1){50}}
				\put(2,29){$\frac{\sigma_{a}\,x}{\sqrt{3/2}}$}
				\put(40,18){$\frac{\sigma_{a}\,x}{\sqrt{3/2}}$}
				\put(15,43){$\frac{\sigma_{b}\,x}{\sqrt{3/2}}$}
				\put(27,4){$\frac{\sigma_{b}\,x}{\sqrt{3/2}}$}
				\put(32.5,36){$\frac{\sigma_{a}+\sigma_{b}}{\sqrt{2}}\,x$}
				\put(7.5,11){$\frac{\sigma_{a}+\sigma_{b}}{\sqrt{2}}\,x$}
				\put(0,45){\rotatebox{-45}{$\sqrt{3}\,x(1-x)$}}
				\put(33,12){\rotatebox{-45}{$\sqrt{3}\,x(1-x)$}}
			\end{picture}
		\end{tabular}
	\end{center}
	\caption{Asymptotic behaviour for \emph{large} $\sigma_{a},\sigma_{b}$ of the average of a single KPZ height (left) and of the combination of two KPZ fields with opposite asymmetry used in the letter for the open spin chain, with usual KPZ normalization (centre) and normalized as the magnetization $m(x,t)$ (right ; this normalization introduces artificial discontinuities when $\sigma_{a}=0$ or $\sigma_{b}=0$).}
	\label{fig phase diag}
\end{figure}

Large values of $\sigma_{a}$ and $\sigma_{b}$ give
\begin{equation}
	\Var(h_{\rm st}^{\sigma_{a},\sigma_{b}}(1))\simeq\left\{
	\begin{array}{cl}
		1/2 & \text{if}\;\sigma_{a}<0\;\text{and}\;\sigma_{b}>0\\
		1 & \text{if}\;(\sigma_{a}>0\;\text{or}\;\sigma_{b}<0)\;\text{and}\;\sigma_{a}+\sigma_{b}\neq0\\
		\frac{\sigma_{a}^{2}}{3}+\frac{4}{3} & \text{if}\;\sigma_{a}=-\sigma_{b}>0
	\end{array}
	\right.\;,
\end{equation}
see figure~\ref{fig Var h}, which leads to the map
\begin{equation}
	(m_{a},m_{b})\simeq\left\{
	\begin{array}{cl}
		(\sqrt{2}\,\sigma_{a},\sqrt{2}\,\sigma_{b}) & \text{if}\;\sigma_{a}\,\sigma_{b}>0\\
		(\frac{2\sqrt{2}\,\sigma_{a}}{\sqrt{3}},\frac{2\sqrt{2}\,\sigma_{b}}{\sqrt{3}}) & \text{if}\;\sigma_{a}\,\sigma_{b}<0\;\text{and}\;\sigma_{a}+\sigma_{b}\neq0\\
		(2\sqrt{3}\sign(\sigma_{a}),2\sqrt{3}\sign(\sigma_{b})) & \text{if}\;\sigma_{a}+\sigma_{b}=0
	\end{array}
	\right.\;,
\end{equation}
between boundary densities and magnetizations. More precisely, near the point $\sigma_{a}=-\sigma_{b}>0$ where $\Var(h_{\rm st}^{\sigma_{a},\sigma_{b}}(1))$ diverges, corresponding to the limit toward the coexistence line for exclusion processes [\citeSD], one has
\begin{equation}
	\Var(h_{\rm st}^{\sigma_{a},\sigma_{b}}(1))\simeq\Big(\frac{1}{(\sigma_{a}+\sigma_{b})^{2}}-\frac{1}{\sinh(\sigma_{a}+\sigma_{b})^{2}}\Big)\sigma_{a}^{2}
\end{equation}
with $\sigma_{a}+\sigma_{b}$ kept fixed when $\sigma_{a}\to+\infty$. The density profile behaves as
\begin{equation}
	\langle h_{\rm st}^{\sigma_{a},\sigma_{b}}(x)\rangle\simeq\left\{
	\begin{array}{cl}
		-2\sqrt{x(1-x)/\pi} & \text{if}\;\sigma_{a}<0\;\text{and}\;\sigma_{b}>0\\
		\sigma_{a}\,x & \text{if}\;\sigma_{a}>0\;\text{and}\;\sigma_{a}+\sigma_{b}>0\\
		\sigma_{b}\,x & \text{if}\;\sigma_{b}<0\;\text{and}\;\sigma_{a}+\sigma_{b}<0\\
		\sigma_{a}\,x(1-x) & \text{if}\;\sigma_{a}=-\sigma_{b}>0
	\end{array}
	\right.\;.
\end{equation}
see figure~\ref{fig phase diag}.

\clearpage
\section{Stationary two-point function for KPZ in finite volume}

\subsection{Exact solution for periodic boundaries}
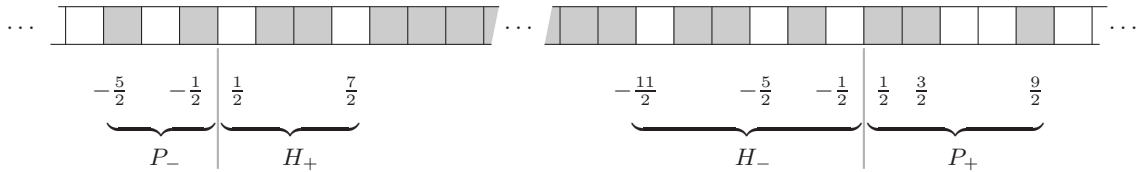
\begin{figure}[h]
	\begin{center}
		\begin{picture}(160,20)(9,0)
			\put(25,15){\color[rgb]{0.8,0.8,0.8}\polygon*(0,0)(5,0)(5,5)(0,5)}
			\put(35,15){\color[rgb]{0.8,0.8,0.8}\polygon*(0,0)(5,0)(5,5)(0,5)}
			\put(45,15){\color[rgb]{0.8,0.8,0.8}\polygon*(0,0)(5,0)(5,5)(0,5)}
			\put(50,15){\color[rgb]{0.8,0.8,0.8}\polygon*(0,0)(5,0)(5,5)(0,5)}
			\put(60,15){\color[rgb]{0.8,0.8,0.8}\polygon*(0,0)(15,0)(15,5)(0,5)}
			\put(75,15){\color[rgb]{0.8,0.8,0.8}\polygon*(0,0)(1,0)(2,5)(0,5)}
			\put(80,15){\color[rgb]{0.8,0.8,0.8}\polygon*(3,0)(5,0)(5,5)(4,5)}
			\put(85,15){\color[rgb]{0.8,0.8,0.8}\polygon*(0,0)(10,0)(10,5)(0,5)}
			\put(100,15){\color[rgb]{0.8,0.8,0.8}\polygon*(0,0)(5,0)(5,5)(0,5)}
			\put(105,15){\color[rgb]{0.8,0.8,0.8}\polygon*(0,0)(5,0)(5,5)(0,5)}
			\put(115,15){\color[rgb]{0.8,0.8,0.8}\polygon*(0,0)(5,0)(5,5)(0,5)}
			\put(125,15){\color[rgb]{0.8,0.8,0.8}\polygon*(0,0)(5,0)(5,5)(0,5)}
			\put(130,15){\color[rgb]{0.8,0.8,0.8}\polygon*(0,0)(5,0)(5,5)(0,5)}
			\put(145,15){\color[rgb]{0.8,0.8,0.8}\polygon*(0,0)(5,0)(5,5)(0,5)}
			\put(12.2,17){\small\ldots}
			\multiput(18,15)(1,5){2}{\line(1,0){58}}\multiput(20,15)(5,0){12}{\line(0,1){5}}
			\put(77.7,17){\small\ldots}
			\multiput(83,15)(1,5){2}{\line(1,0){73}}\multiput(85,15)(5,0){15}{\line(0,1){5}}
			\put(157.2,17){\small\ldots}
			\put(40,14.5){\thicklines\color[rgb]{0.7,0.7,0.7}\line(0,-1){16}}
			\put(125,14.5){\thicklines\color[rgb]{0.7,0.7,0.7}\line(0,-1){16}}
			\put(23.5,8){$-\frac{5}{2}$}
			\put(33.5,8){$-\frac{1}{2}$}
			\put(41.5,8){$\frac{1}{2}$}
			\put(56.5,8){$\frac{7}{2}$}
			\put(92,8){$-\frac{11}{2}$}
			\put(108.5,8){$-\frac{5}{2}$}
			\put(118.5,8){$-\frac{1}{2}$}
			\put(126.5,8){$\frac{1}{2}$}
			\put(131.5,8){$\frac{3}{2}$}
			\put(146.5,8){$\frac{9}{2}$}
			\put(25.5,5){$\underbrace{\hspace{14\unitlength}}$}
			\put(40.5,5){$\underbrace{\hspace{18\unitlength}}$}
			\put(94.5,5){$\underbrace{\hspace{30\unitlength}}$}
			\put(125.5,5){$\underbrace{\hspace{23\unitlength}}$}
			\put(31,-1){$P_{-}$}
			\put(48.5,-1){$H_{+}$}
			\put(108,-1){$H_{-}$}
			\put(136,-1){$P_{+}$}
		\end{picture}
	\end{center}\vspace{-5mm}
	\caption{Particle-hole excitations contributing to the stationary two-point function $c_{2}(x,t)$ for KPZ with periodic boundaries.}
	\label{fig particle-hole}
\end{figure}

The stationary two-point function $c_{2}(x,t)$ of the KPZ density field with periodic boundary condition has the following exact Fourier representation [\citePeriodicKPZ] (an alternative expression involving an integral of a Fredholm determinant is also available [\citeZL])
\begin{equation}
	\label{c2(x,t)}
	c_{2}(x,t)=-(2\pi)^{5/2}\sum_{P,H}\ee^{2\ii\pi Kx}\,K^{2}(\ii\pi/2)^{2m^{2}}V(P)^{2}\,V(H)^{2}\,\frac{\ee^{t\chi(\nu)+\int\hspace{-2.6mm}-_{-\infty}^{\nu}\dd v\,\chi''(v)^{2}}}{\ee^{\nu}\chi''(\nu)}\;,
\end{equation}
with $V(S)=\prod_{a>b\in S}(a-b)$.
The summation is over finite sets of half-integers $P\subset\mathbb{Z}+1/2$ and $H\subset\mathbb{Z}+1/2$ with the same cardinal $m=|P|=|H|$, representing the momenta of quasi-particles and holes counted from the Fermi momentum, see figure~\ref{fig particle-hole}. The total momentum associated with the eigenstate is
\begin{equation}
	\label{K[P,H]}
	K=\sum_{a\in P}a+\sum_{a\in H}a\;.
\end{equation}
The sets $P$ and $H$ verify the additional constraint $|P_{+}|=|H_{-}|$ and $|P_{-}|=|H_{+}|$, where the subscript $_{+}$ ($_{-}$) stand for keeping only the positive (negative) element. Physically, this means that particle-hole excitations occur independently on both sides of the Fermi sea, in contrast with the standard Luttinger liquid picture of one-dimensional quantum systems where quasiparticles can be sent from one side of the Fermi sea to the other. Such processes are eliminated by the anomalous KPZ scaling, and only the central region with momentum $|K|$ much smaller than the Fermi momentum contributes here, see [\citeHDR].

The function $\chi$ in (\ref{c2(x,t)}) depends on $P$ and $H$, and the quantity $\nu$ is the unique solution of $\chi'(\nu)=0$. More precisely, the function $\chi$ is defined by
\begin{equation}
	\chi(v)=-\frac{\Li_{5/2}(-\ee^{v})}{\sqrt{2\pi}}+\sum_{a\in P}\frac{\omega_{a}^{3}(v)}{3}+\sum_{a\in H}\frac{\omega_{a}^{3}(v)}{3}\;,
\end{equation}
with the polylogarithm $\Li_{5/2}(z)=\sum_{n=1}^{\infty}\frac{z^{n}}{n^{5/2}}$ for $|z|<1$ analytically continued so that $\Li_{5/2}(-\ee^{v})$ is analytic outside the branch cut $v\in\ii(\mathbb{R}\setminus(-\pi,\pi))$, and with
$\omega_{a}(v)=\sqrt{2\sign(a)\ii}\,\sqrt{2\pi|a|+\sign(a)\ii v}$ where the square root is taken with branch cut $\mathbb{R}^{-}$. The integral of $\chi''(v)^{2}$ in (\ref{c2(x,t)}) is regularized as $\int\hspace{-3.1mm}-_{-\infty}^{\nu}\chi''(v)^{2}=\lim\limits_{\Lambda\to\infty}\big(-2m^{2}\log\Lambda+\int_{-\Lambda}^{\nu}\dd v\,\chi''(v)^{2}\big)$.

The functions $\chi$ for various $P,H$ but with the same $\Delta=(P\cup H)\backslash(P\cap H)$ are recovered from one another by analytic continuation through the branch cut $\ii(\mathbb{R}\setminus(-\pi,\pi))$, and $c_{2}(x,t)$ can instead be seen as a sum over Riemann surfaces $\mathcal{R}_{\Delta}$ where the elements of $\Delta=(P\cup H)\backslash(P\cap H)$ specify branch points that have been removed from the polylogarithm, see [\citeRS]. All the sheets $(P,H)$ of $\mathcal{R}_{\Delta}$ contribute the same spatial dependency
\begin{equation}
	\sum_{{A\subset\Delta}\atop{2|A|=|\Delta|}}\ee^{2\ii\pi x\big(\sum\limits_{a\in A}a-\sum\limits_{a\in\Delta\backslash A}a\big)}\,V(A)^{2}\,V(\Delta\backslash A)^{2}
\end{equation}
to $c_{2}(x,t)$.

\begin{table}
	\begin{center}
		\vspace{3mm}
		$\begin{array}{|c|c|r|r|r|r|}\hline
			k & n & \tau_{n,k} & \omega_{n,k} & \varphi_{n,k} & b_{n,k}\\\hline\hline
			1 & 1 & 0.076814 & 0 & 0 & 1.086880 \\\hline
			1 & 2 & 0.017956 & 10.02330 & -1.912611 & 0.204032 \\\hline
			1 & 3 & 0.009324 & 9.133798 & -1.573129 & 0.135161 \\\hline
			1 & 4 & 0.009102 & 34.87794 & 0.836336 & 0.068712 \\\hline
			1 & 5 & 0.009055 & 16.50288 & -2.257356 & 0.089518 \\\hline\hline
			2 & 1 & 0.029089 & 10.87324 & 0.783930 & 1.688000 \\\hline
			2 & 2 & 0.012355 & 0 & 0 & -0.416212 \\\hline
			2 & 3 & 0.011899 & 26.64389 & -1.013067 & 0.513196 \\\hline\hline
			3 & 1 & 0.017275 & 0 & 0 & 1.211412 \\\hline
			3 & 2 & 0.016329 & 28.10454 & 1.516568 & 1.585887 \\\hline
			3 & 3 & 0.009453 & 0 & 0 & 0.082243 \\\hline
			3 & 4 & 0.009055 & 16.50288 & -2.257356 & 1.812730 \\\hline\hline
			4 & 1 & 0.012355 & 0 & 0 & -0.104053 \\\hline
			4 & 2 & 0.011605 & 17.14676 & 0.663978 & 3.115857 \\\hline
			4 & 3 & 0.010801 & 50.33536 & 2.207111 & 1.610726 \\\hline\hline
			5 & 1 & 0.009055 & 16.50288 & -2.257356 & 0.559485 \\\hline
		\end{array}$
		\vspace{-3mm}
	\end{center}
	\caption{Numerical values of the coefficients entering the Fourier modes $a_{k}(t)$ of the stationary two-point function $c_{2}(x,t)$ for KPZ with periodic boundary conditions. All the modes $k,n$ with relaxation time $\tau_{n,k}>0.009$ are shown.}
	\label{table coeffs c2 P}
\end{table}

The contribution of each $P,H$ to $c_{2}(x,t)$ involves in general complex numbers. Putting together the contributions of the particle-hole excitations $(P,H)$, $(-P,-H)$, $(H,P)$, $(-H,-P)$ gives an expansion with only real coefficients. This leads in particular for the coefficients $a_{k}(t)$ in the Fourier expansion
\begin{equation}
	c_{2}(x,t)=2\sum_{k=1}^{\infty}a_{k}(t)\cos(2\pi kx)
\end{equation}
to a representation as an infinite sum $a_{k}(t)=\sum_{n=1}^{\infty}b_{n,k}\,\ee^{-t/\tau_{n,k}}\cos(\omega_{n,k}t+\varphi_{n,k})$ of exponentially decaying terms, with $n$ indexing the particle-hole excitations $(P,H)$ with total momentum $K=\pm k$, and real valued coefficients $\tau_{n,k}$, $\omega_{n,k}$, $\varphi_{n,k}$, $b_{n,k}$, see table~\ref{table coeffs c2 P} for numerical values of the first coefficients. At any finite time $t$, the infinite sum over $n$ can be truncated with good accuracy, but shorter times require more terms. In the letter, working with a numerical evaluation of up to $\approx35000$ terms allows us to consider times $t>0.002$. The supplemental material contains csv files with numerical data for the first terms in the expansion of $c_{2}(x,t)$ corresponding to particle-hole excitations up to ``size'' $Q=10$, $Q=20$ and $Q=30$, where
\begin{equation}
	Q=\sum_{a\in P}|a|+\sum_{a\in H}|a|\;.
\end{equation}

\subsection{TASEP numerics for open boundaries}
While no exact expression is known, accurate numerical values can be obtained from simulations of the totally asymmetric simple exclusion process (TASEP), a Markov process describing classical hard-core particles moving in the same direction on a one-dimensional lattice, and whose scaling limit to KPZ has been proved rigorously. Boundary densities $\sigma_{a}$, $\sigma_{b}$ are related to injection and ejection rates of particles at the boundaries, see e.g. [\citeHDR]. Numerical simulations of TASEP with $L=255$ sites were used.

\clearpage
\section{Numerical simulations}
For the KPLL model, in order to prevent the accumulation of numerical errors, we set the norms to $1$ at each time step rather than using the exact denominator in (1).

For the Ishimori chain, numerical simulations are more involved than for the KPLL model, as they consist in solving non-linear ordinary differential equations of large dimension with a singularity when two neighbouring spins point in opposite directions. For the small number of spins considered in this letter, however, relatively unsophisticated methods (i.e. Mathematica's NDSolve function [\citeWolfram] with default parameters) were good enough. Because of the singularity of the differential equations, the numerical resolution fails in about $0.02\%$ and $0.06\%$ of the cases for $L=15$ and $L=31$ spins, and we simply throw away the corresponding data points.

\clearpage
\section{Magnetization profiles with open boundaries}
\begin{figure}[b]
	\includegraphics[width=160mm]{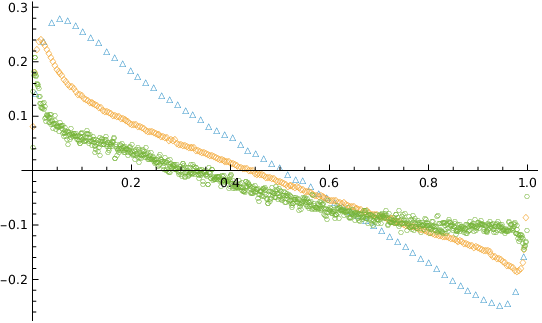}
	\begin{picture}(0,0)
		\put(-4,48){$x$}
		\put(-25,80){
			\begin{tabular}{clc}
				$L=63$ &\hspace*{-2.7mm}& \begin{tabular}{c}\includegraphics[width=2.7mm]{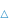}\end{tabular}\\[-2.5mm]
				$L=255$ &\hspace*{-2.7mm}& \begin{tabular}{c}\includegraphics[width=2.7mm]{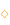}\end{tabular}\\[-3mm]
				$L=1023$ &\hspace*{-2.7mm}& \begin{tabular}{c}\includegraphics[width=2.5mm]{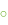}\end{tabular}
			\end{tabular}
		}
	\end{picture}
	\caption{Difference between the KPLL simulations and the exact KPZ result for the stationary magnetization profile with $s_{a}=1$, $s_{b}=-0.5$, plotted here directly for the stationary magnetization profile rather than for the height profile as in figure~2b of the letter.}
	\label{fig error magn profile}
\end{figure}

\begin{figure}
	\hspace*{-19mm}
	\begin{tabular}{lllll}
		\includegraphics[width=40mm]{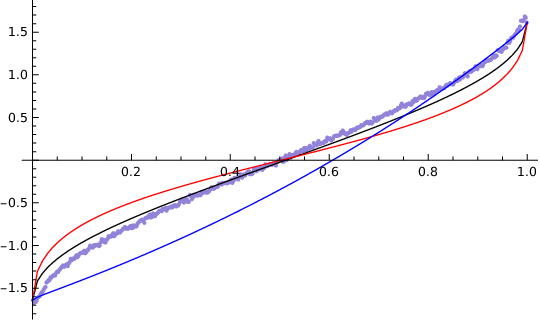} &
		\includegraphics[width=40mm]{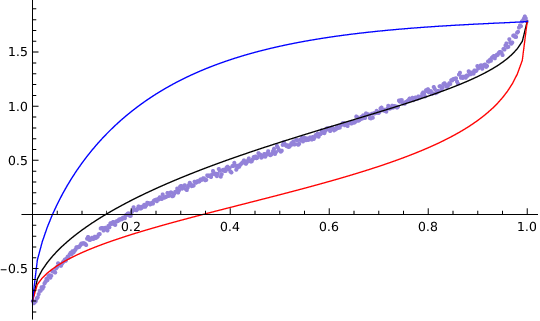} &
		\includegraphics[width=40mm]{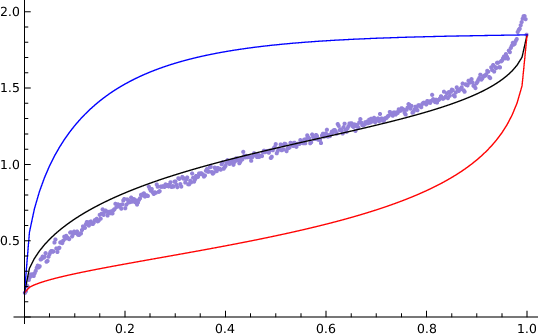} &
		\includegraphics[width=40mm]{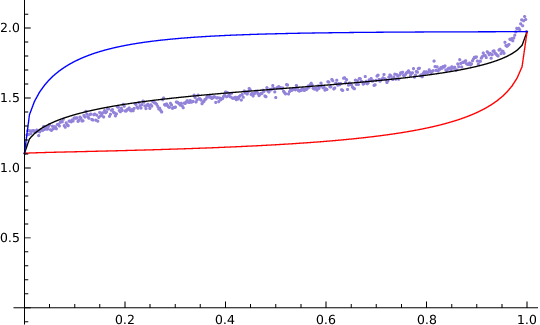} &
		\includegraphics[width=40mm]{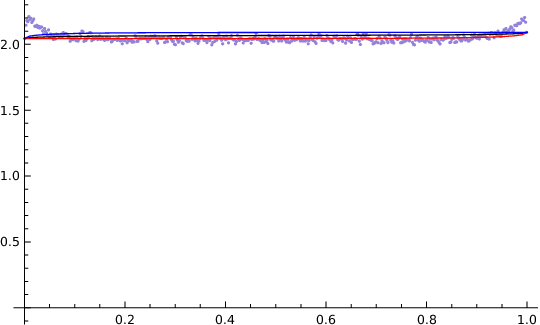}\\
		\includegraphics[width=40mm]{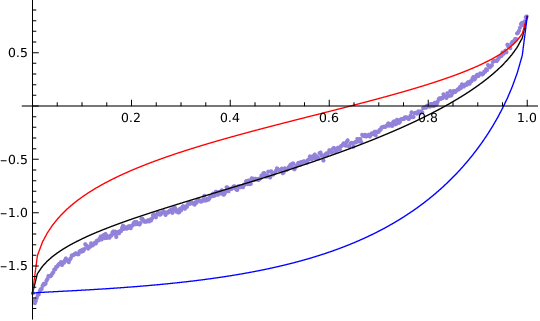} &
		\includegraphics[width=40mm]{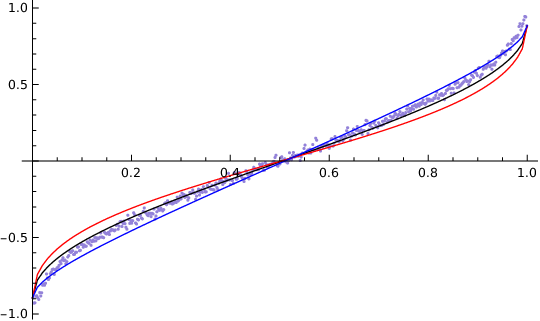} &
		\includegraphics[width=40mm]{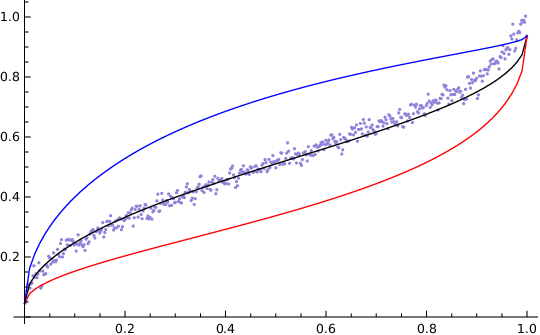} &
		\includegraphics[width=40mm]{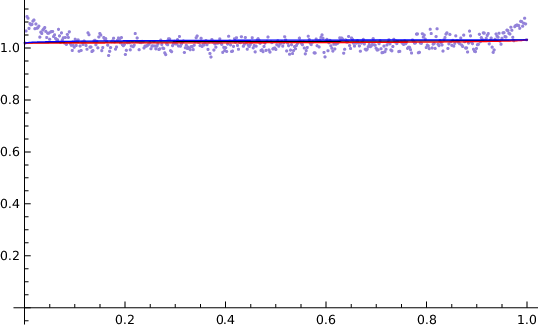} &
		\includegraphics[width=40mm]{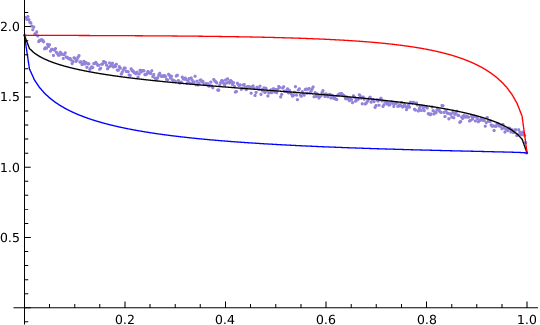}\\
		\includegraphics[width=40mm]{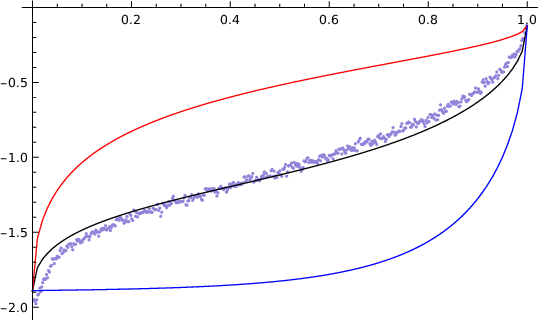} &
		\includegraphics[width=40mm]{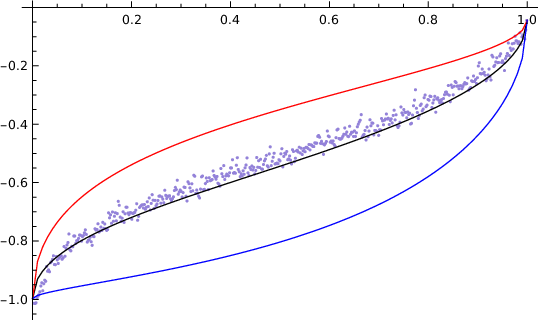} &
		\includegraphics[width=40mm]{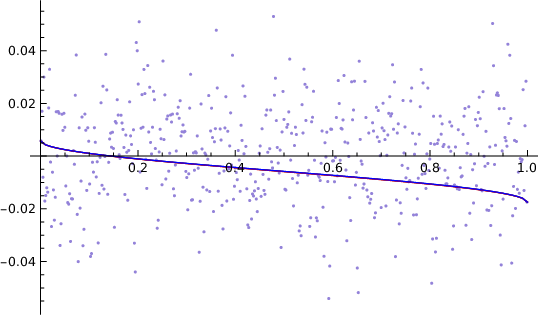} &
		\includegraphics[width=40mm]{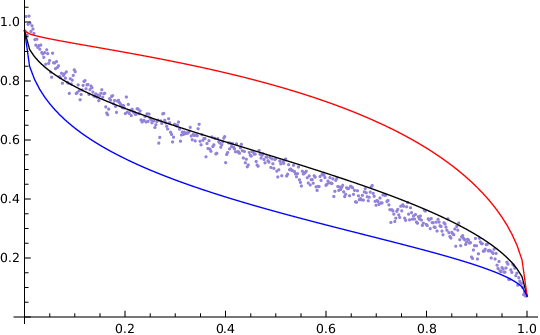} &
		\includegraphics[width=40mm]{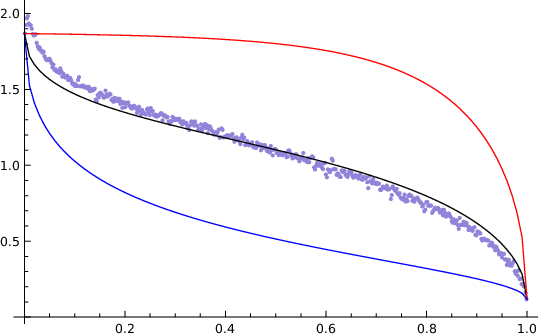}\\
		\includegraphics[width=40mm]{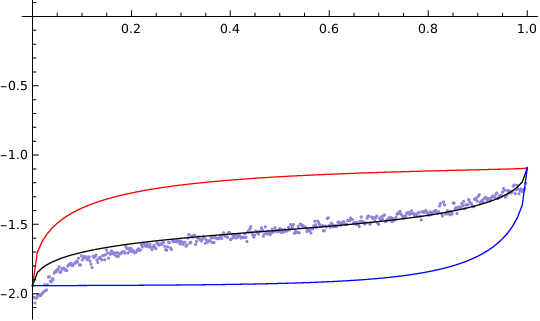} &
		\includegraphics[width=40mm]{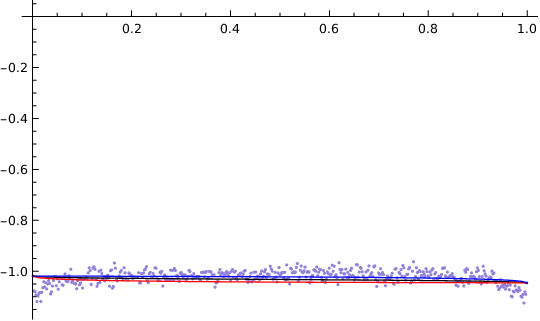} &
		\includegraphics[width=40mm]{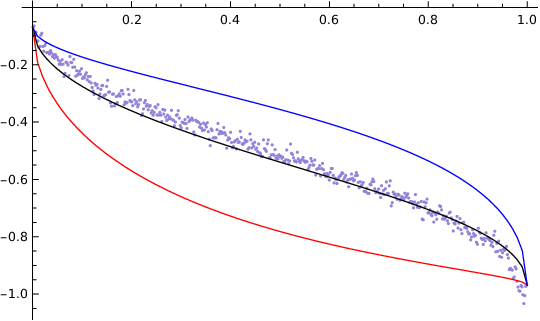} &
		\includegraphics[width=40mm]{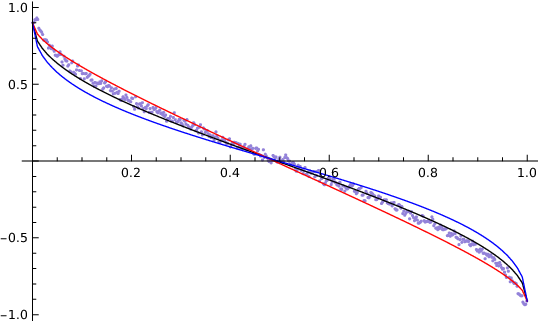} &
		\includegraphics[width=40mm]{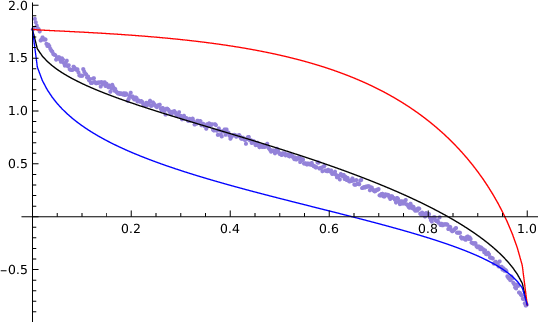}\\
		\includegraphics[width=40mm]{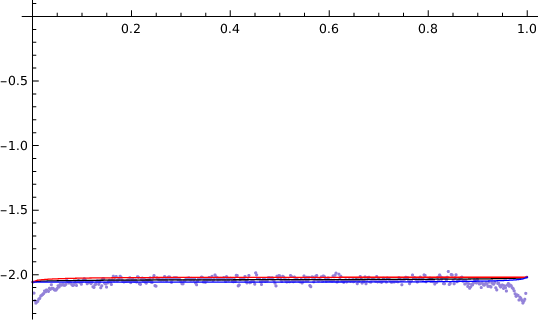} &
		\includegraphics[width=40mm]{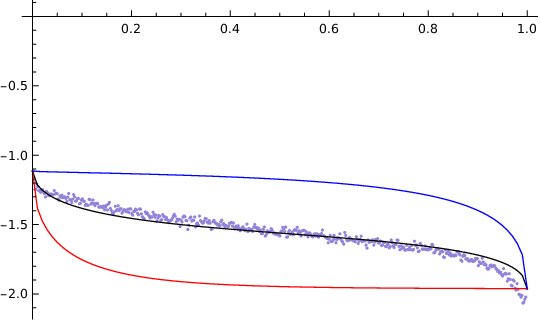} &
		\includegraphics[width=40mm]{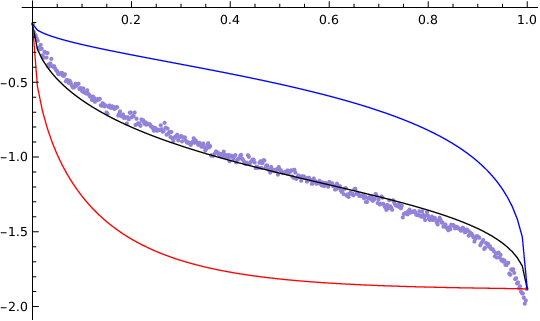} &
		\includegraphics[width=40mm]{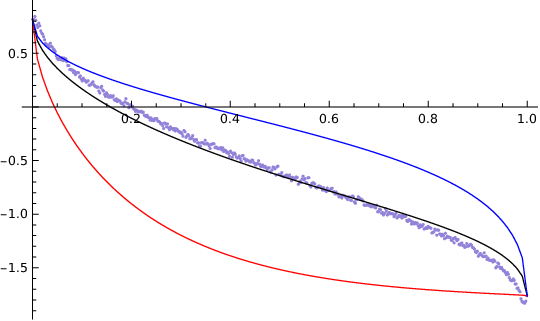} &
		\includegraphics[width=40mm]{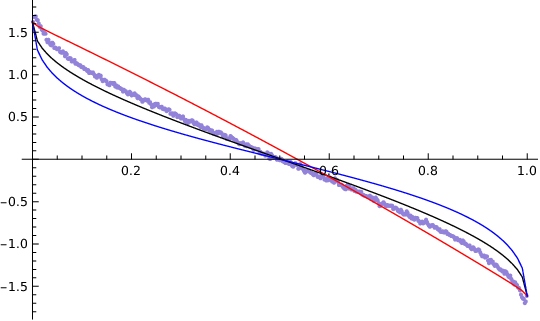}
	\end{tabular}
	\caption{Stationary magnetization profile plotted as a function of the position $x$. The dots represent $\langle S_{i=xL}^{\rm z}\rangle/\Var(M)^{1/2}$ obtained from numerical simulations of the KPLL model with $L=511$ spins, $T=10^{5}$ time steps, averaged over $10^{6}$ realizations, with boundary parameters $s_{a}$, $s_{b}$ taking the values $-1,-0.5,0,0.5,1$ respectively horizontally from the left and vertically from the bottom. The solid black line in the middle is the exact KPZ result. The solid red and blue lines below and above correspond to keeping a single KPZ mode with either sign for the asymmetry.}
	\label{fig C1(x) open}
\end{figure}

\begin{figure}
	\hspace*{-19mm}
	\begin{tabular}{lllll}
		\includegraphics[width=40mm]{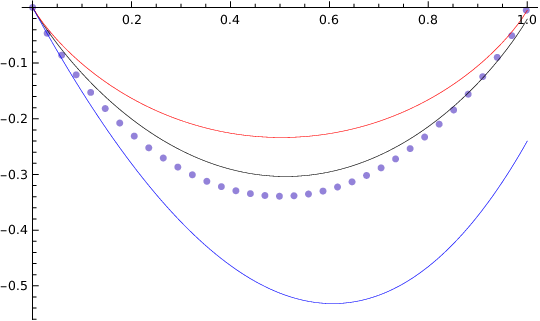} &
		\includegraphics[width=40mm]{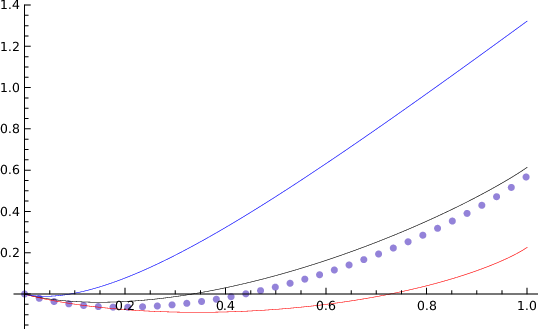} &
		\includegraphics[width=40mm]{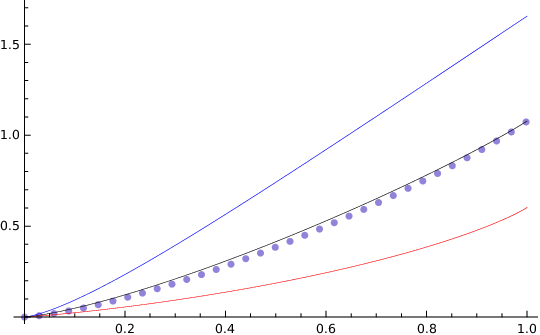} &
		\includegraphics[width=40mm]{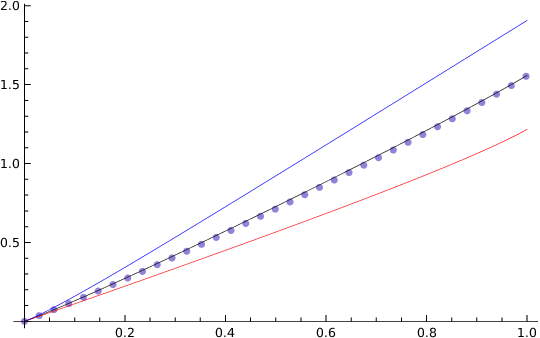} &
		\includegraphics[width=40mm]{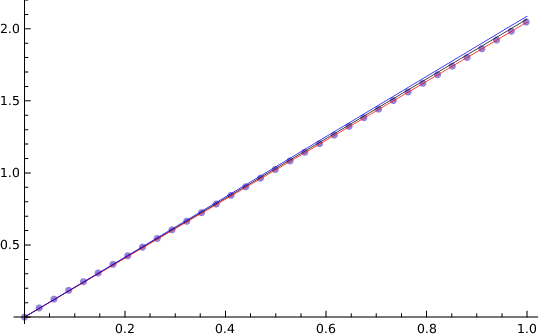}\\
		\includegraphics[width=40mm]{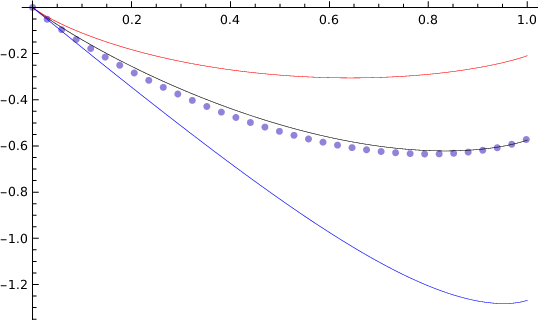} &
		\includegraphics[width=40mm]{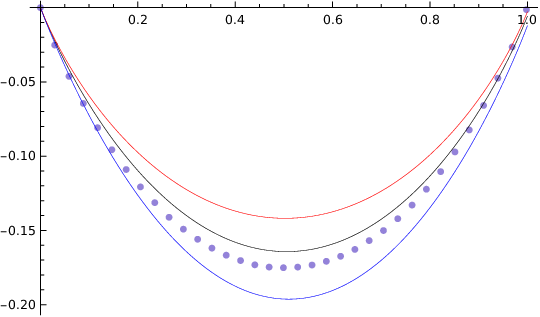} &
		\includegraphics[width=40mm]{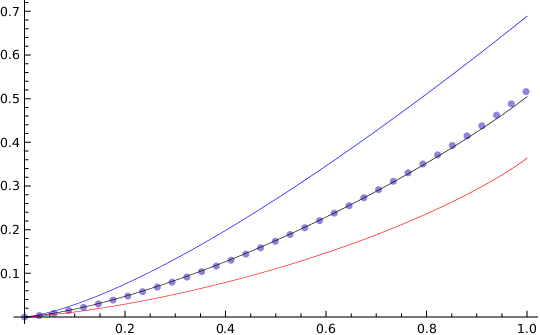} &
		\includegraphics[width=40mm]{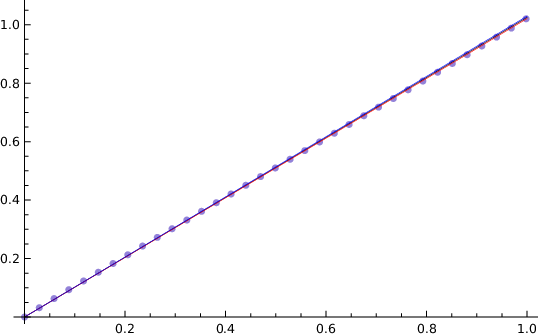} &
		\includegraphics[width=40mm]{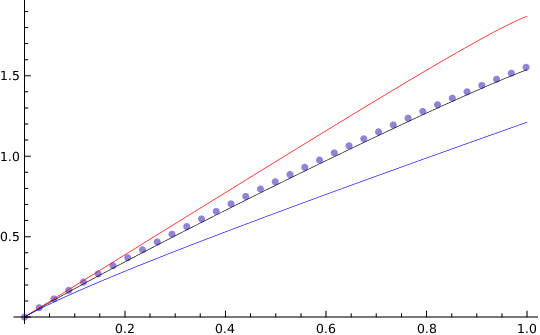}\\
		\includegraphics[width=40mm]{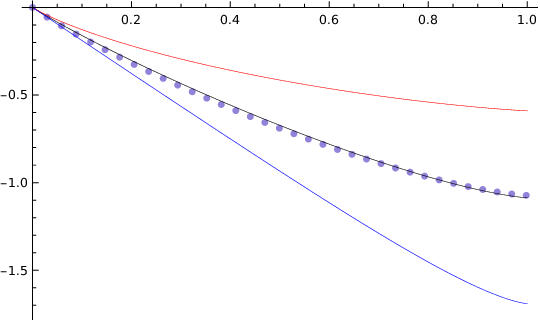} &
		\includegraphics[width=40mm]{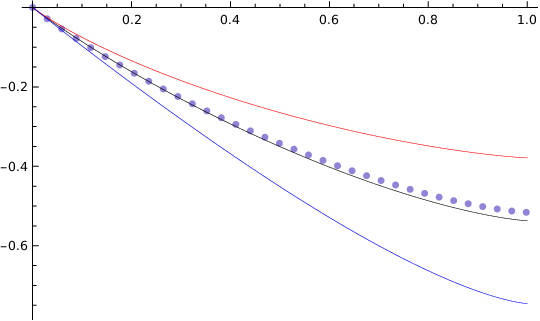} &
		\includegraphics[width=40mm]{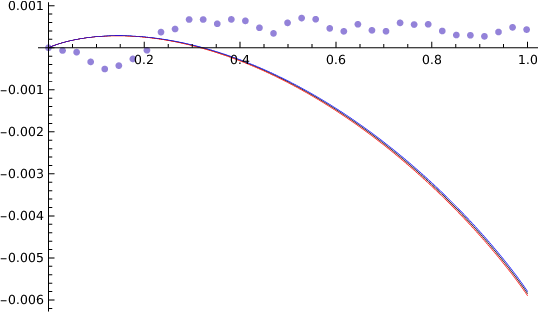} &
		\includegraphics[width=40mm]{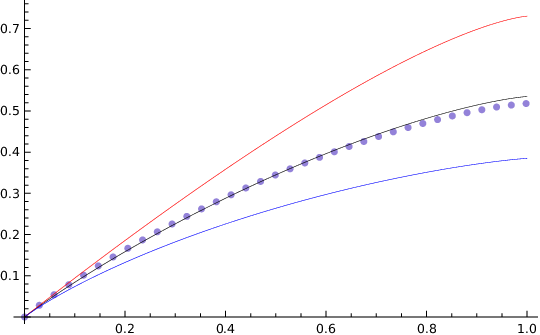} &
		\includegraphics[width=40mm]{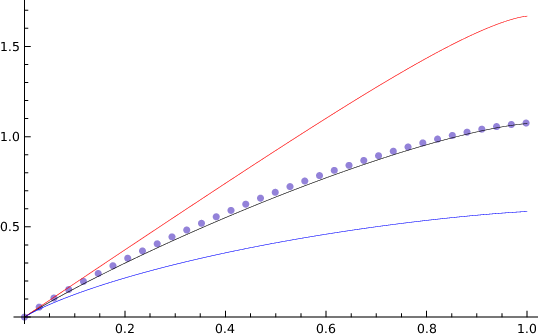}\\
		\includegraphics[width=40mm]{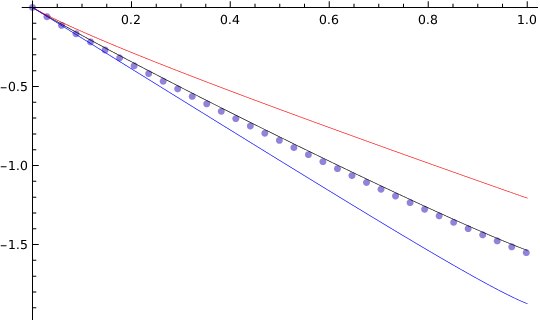} &
		\includegraphics[width=40mm]{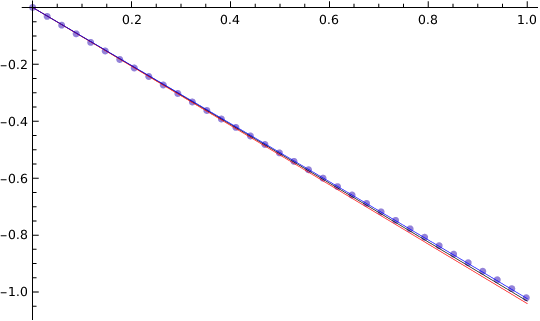} &
		\includegraphics[width=40mm]{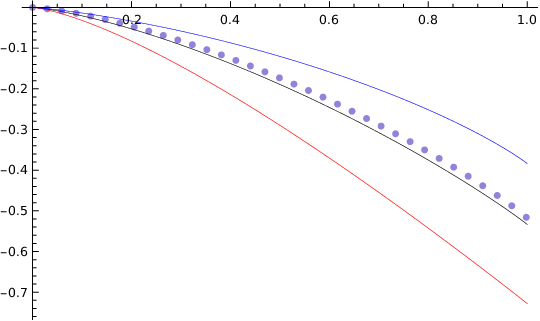} &
		\includegraphics[width=40mm]{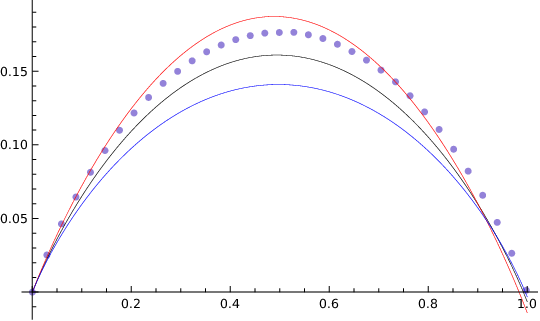} &
		\includegraphics[width=40mm]{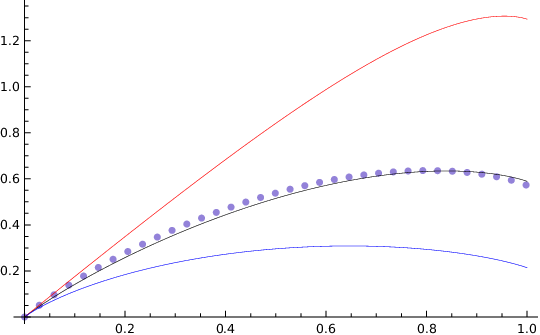}\\
		\includegraphics[width=40mm]{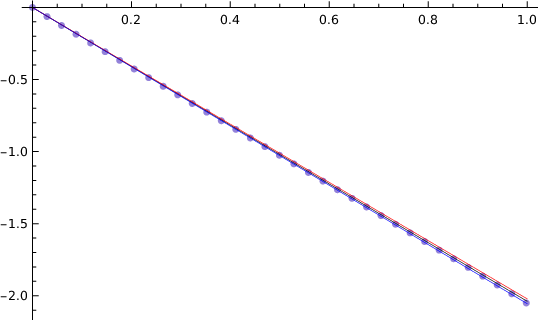} &
		\includegraphics[width=40mm]{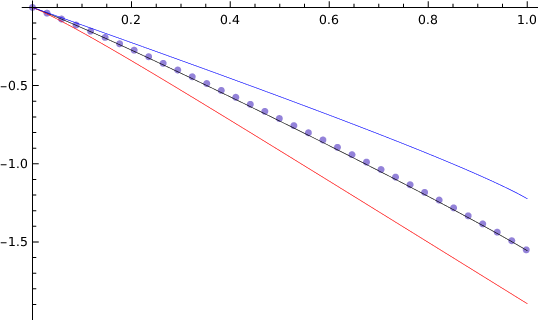} &
		\includegraphics[width=40mm]{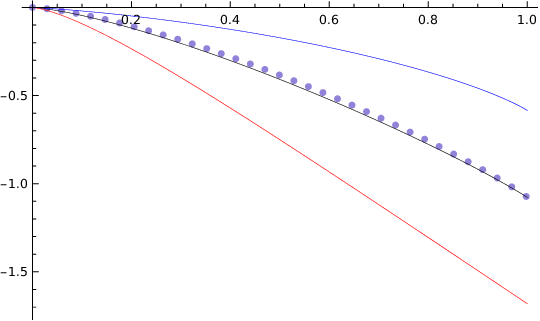} &
		\includegraphics[width=40mm]{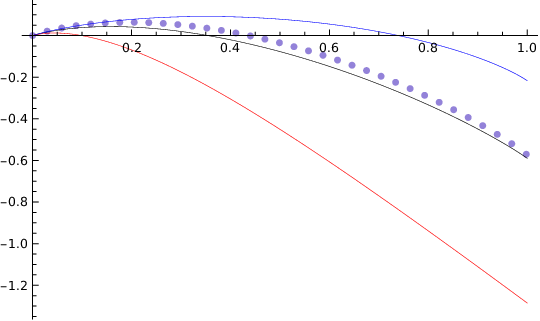} &
		\includegraphics[width=40mm]{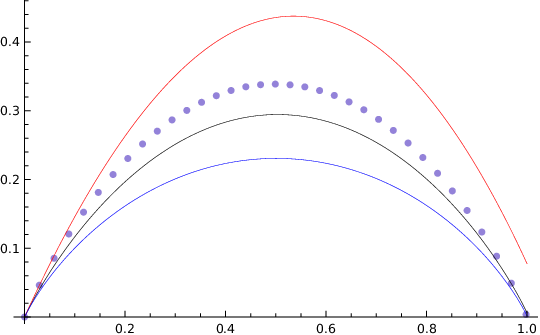}
	\end{tabular}
	\caption{Stationary magnetization profile (integrated over space) plotted as a function of the position $x$. The dots represent $\sum_{i=1}^{xL}\langle S_{i}^{\rm z}\rangle/\Var(M)^{1/2}$ obtained from numerical simulations of the KPLL model with $L=511$ spins, $T=10^{5}$ time steps, averaged over $10^{6}$ realizations, with boundary parameters $s_{a}$, $s_{b}$ taking the values $-1,-0.5,0,0.5,1$ respectively horizontally from the left and vertically from the bottom. The solid black line in the middle is the exact KPZ result. The solid red and blue lines below and above correspond to keeping a single KPZ mode with either sign for the asymmetry.}
	\label{fig int C1(x) open}
\end{figure}

\begin{figure}
	\hspace*{-19mm}
	\begin{tabular}{lllll}
		\includegraphics[width=40mm]{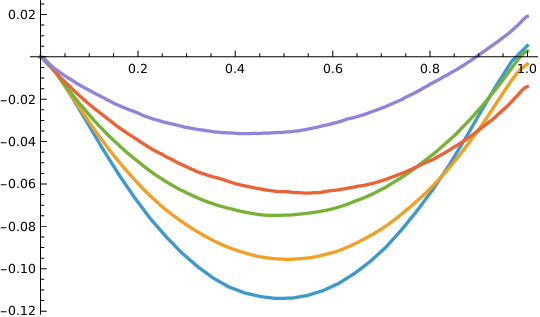} &
		\includegraphics[width=40mm]{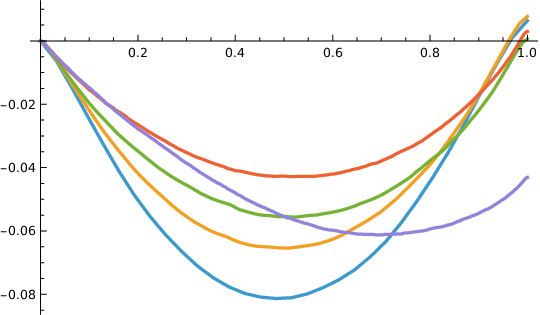} &
		\includegraphics[width=40mm]{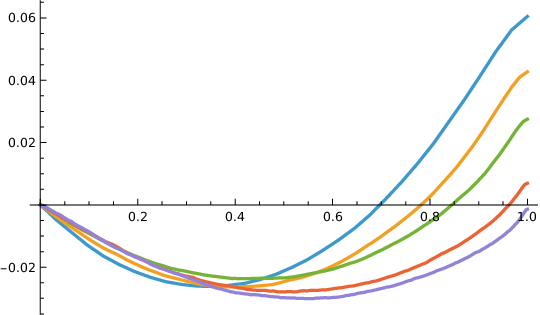} &
		\includegraphics[width=40mm]{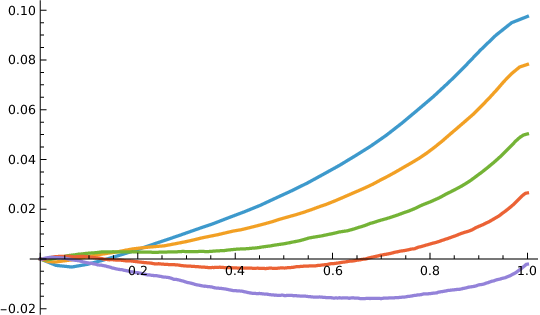} &
		\includegraphics[width=40mm]{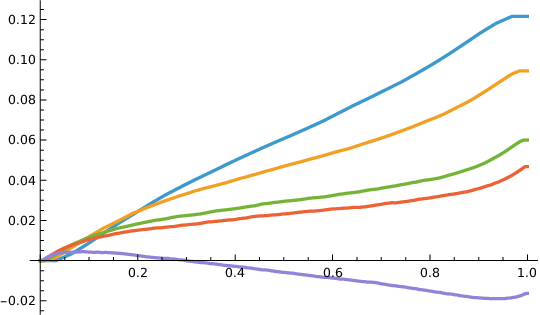}\\
		\includegraphics[width=40mm]{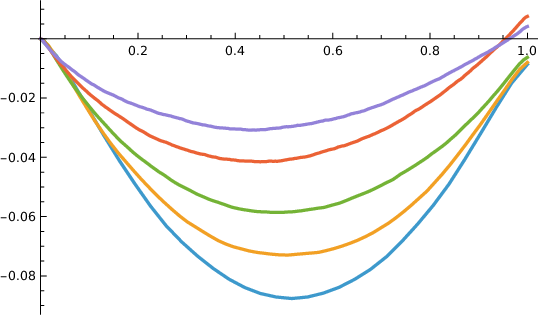} &
		\includegraphics[width=40mm]{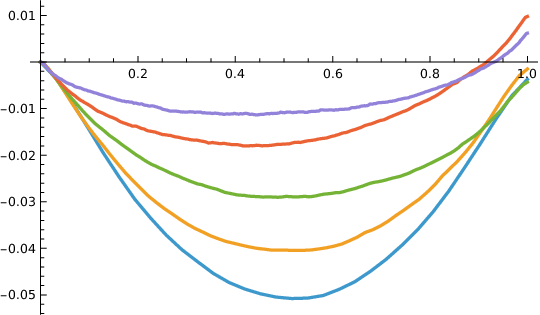} &
		\includegraphics[width=40mm]{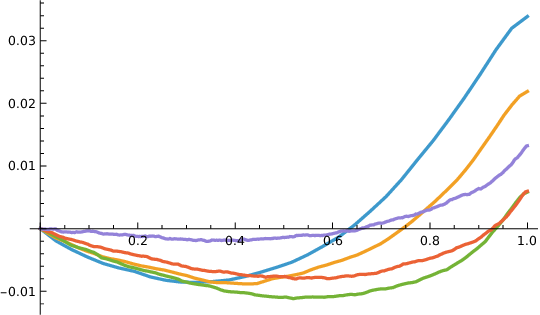} &
		\includegraphics[width=40mm]{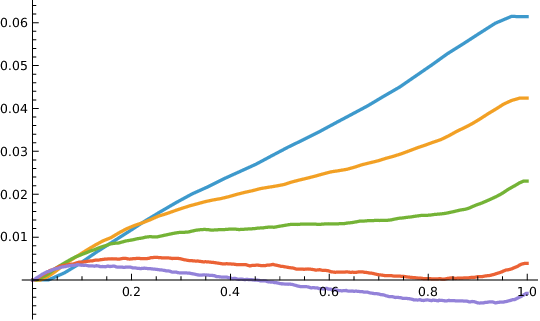} &
		\includegraphics[width=40mm]{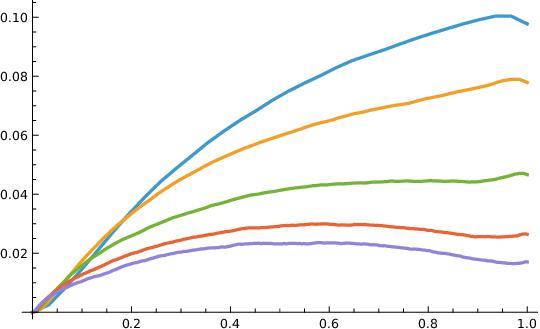}\\
		\includegraphics[width=40mm]{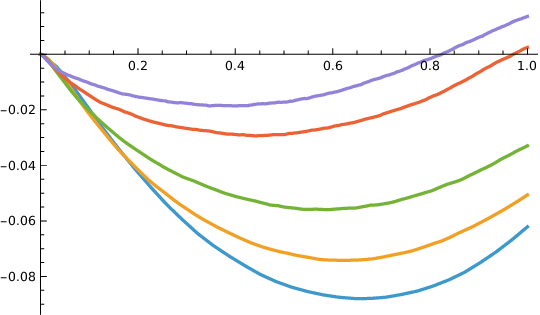} &
		\includegraphics[width=40mm]{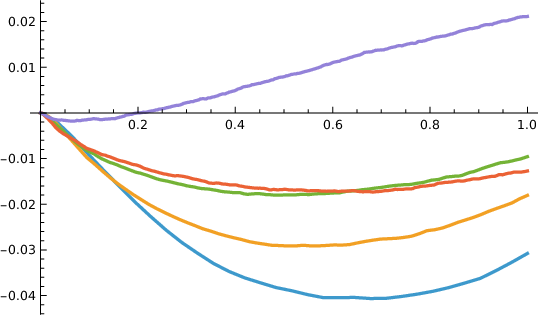} &
		\includegraphics[width=40mm]{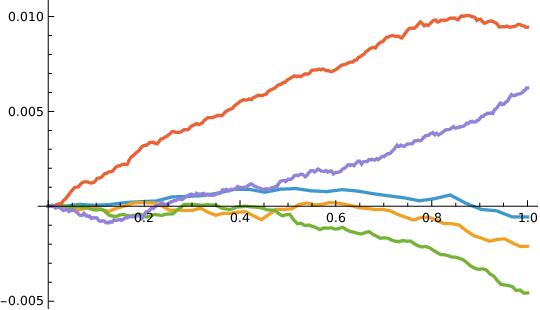} &
		\includegraphics[width=40mm]{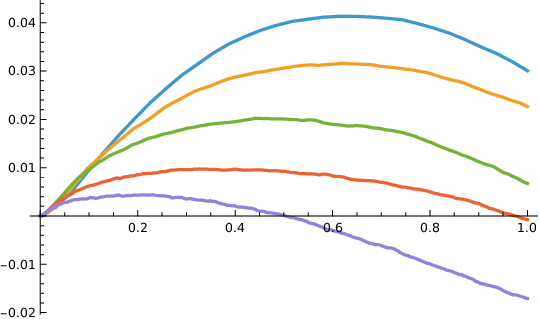} &
		\includegraphics[width=40mm]{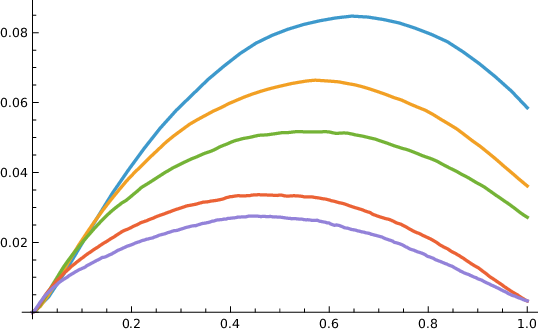}\\
		\includegraphics[width=40mm]{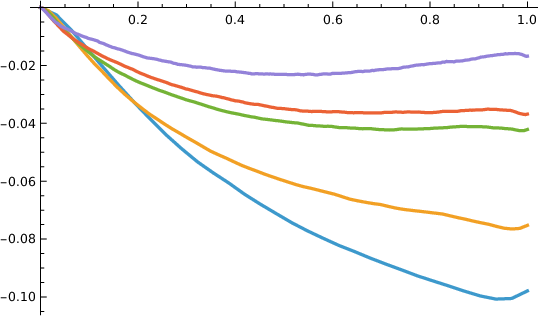} &
		\includegraphics[width=40mm]{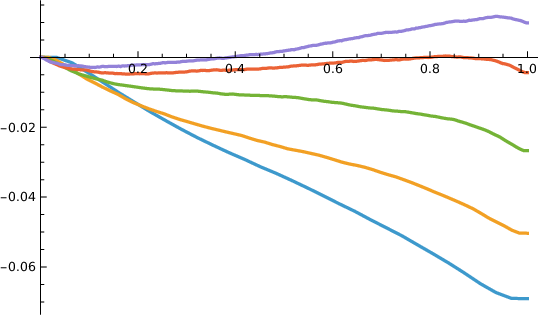} &
		\includegraphics[width=40mm]{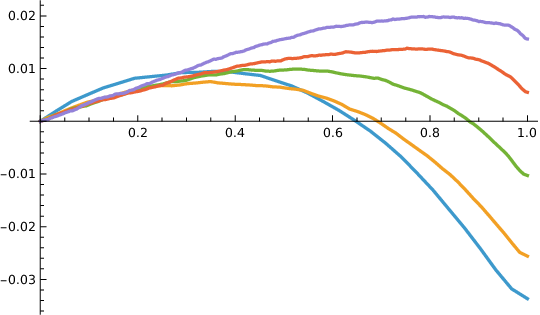} &
		\includegraphics[width=40mm]{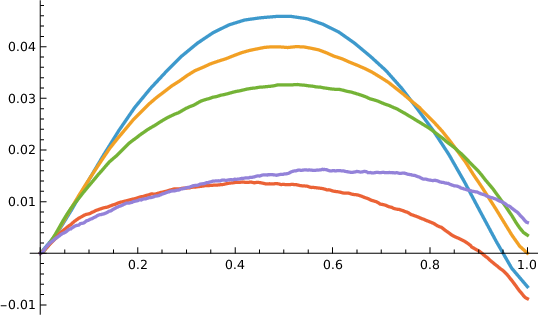} &
		\includegraphics[width=40mm]{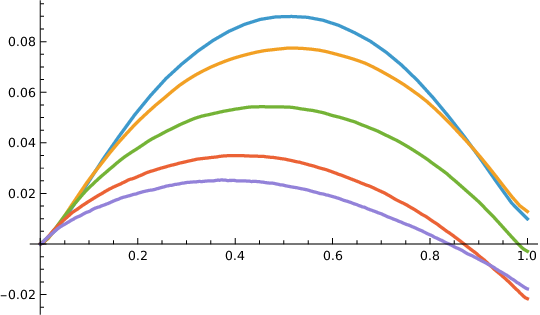}\\
		\includegraphics[width=40mm]{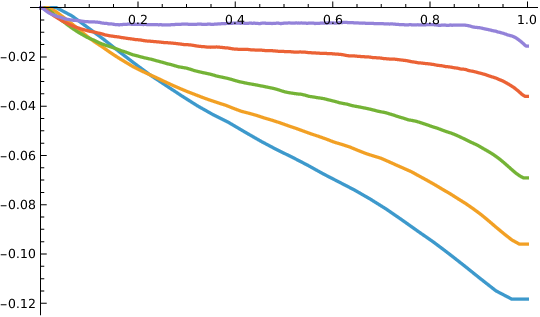} &
		\includegraphics[width=40mm]{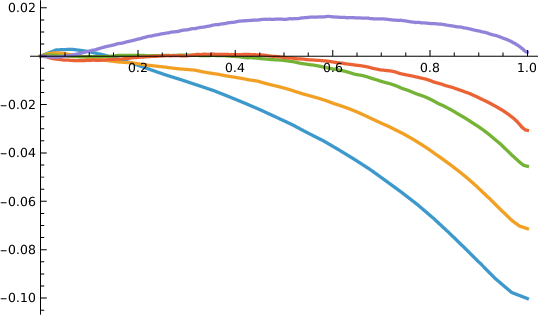} &
		\includegraphics[width=40mm]{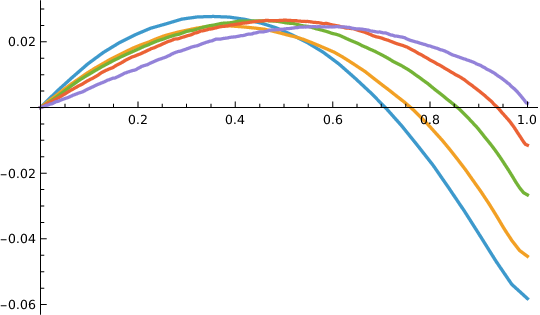} &
		\includegraphics[width=40mm]{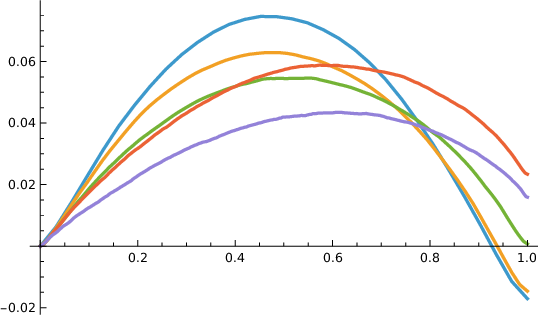} &
		\includegraphics[width=40mm]{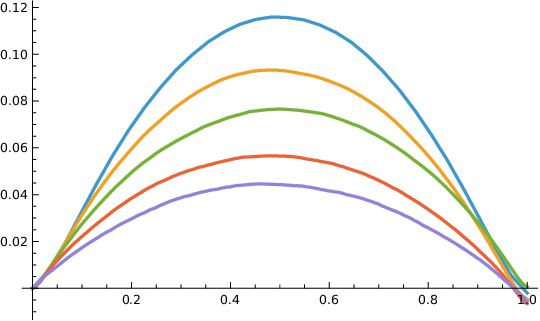}
	\end{tabular}
	\caption{Difference between the KPLL simulation results with $L=31,63,127,255,511$ (in order of decreasing amplitude) and the exact KPZ result (with $\sigma_{a},\sigma_{b}$ adjusted independently in each case), with the same parameters as in figure~\ref{fig C1(x) open}.}
	\label{fig int C1(x) open diff}
\end{figure}

\begin{figure}
	\hspace*{-19mm}
	\begin{tabular}{lllll}
		\includegraphics[width=40mm]{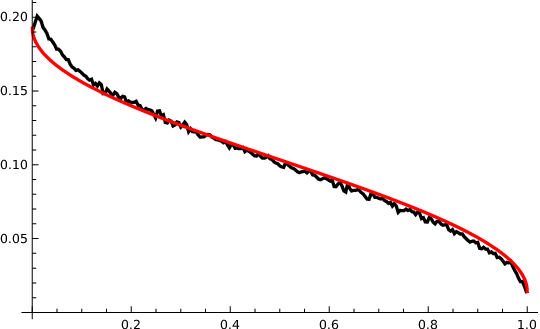} &
		\includegraphics[width=40mm]{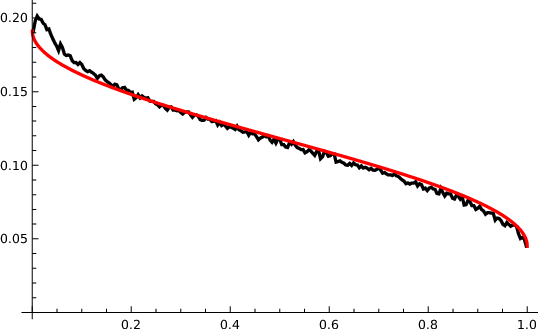} &
		\includegraphics[width=40mm]{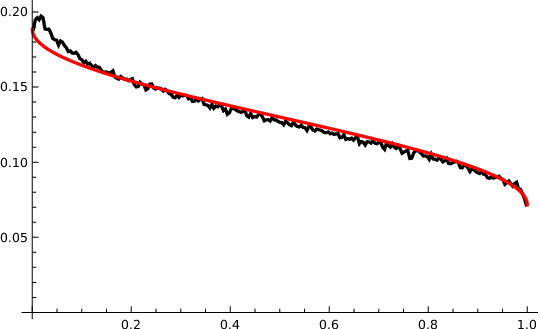} &
		\includegraphics[width=40mm]{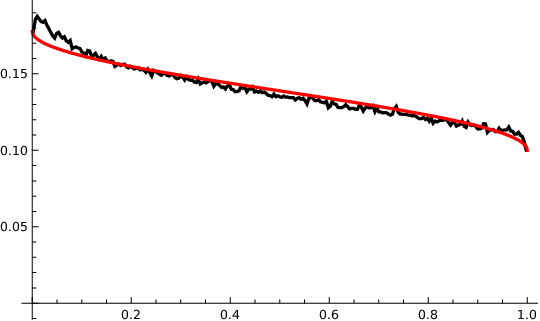} &
		\includegraphics[width=40mm]{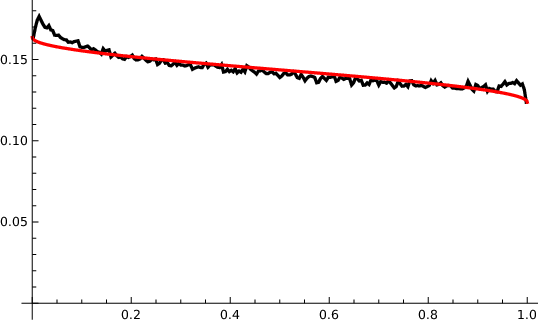}\\
		\includegraphics[width=40mm]{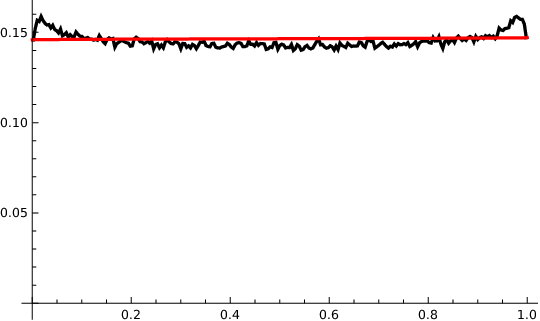} &
		\includegraphics[width=40mm]{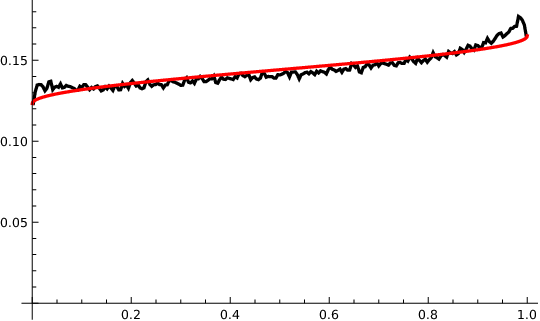} &
		\includegraphics[width=40mm]{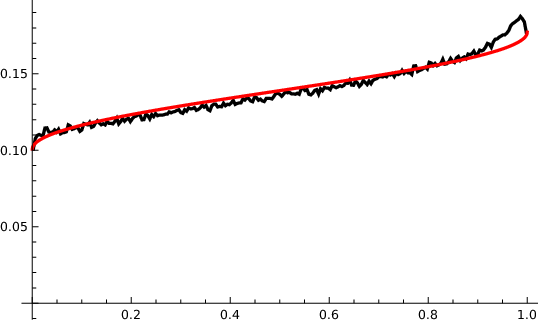} &
		\includegraphics[width=40mm]{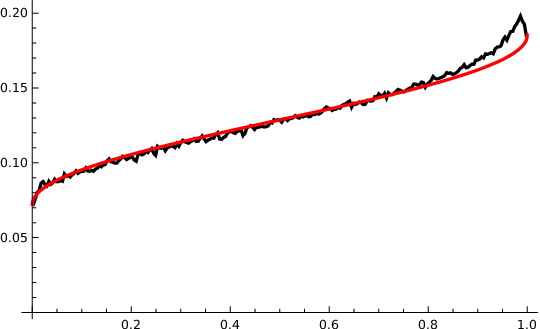} &
		\includegraphics[width=40mm]{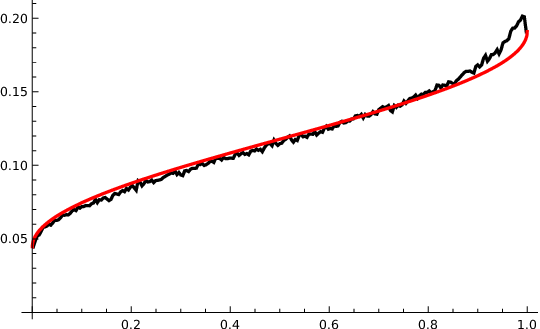}\\
		\includegraphics[width=40mm]{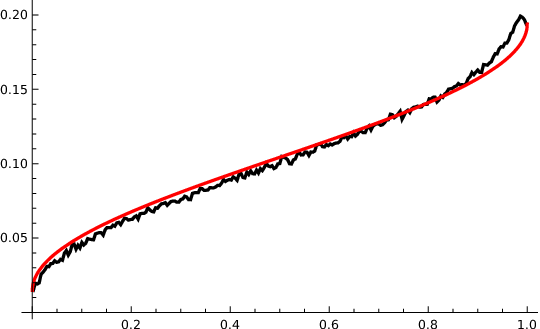} &
		\includegraphics[width=40mm]{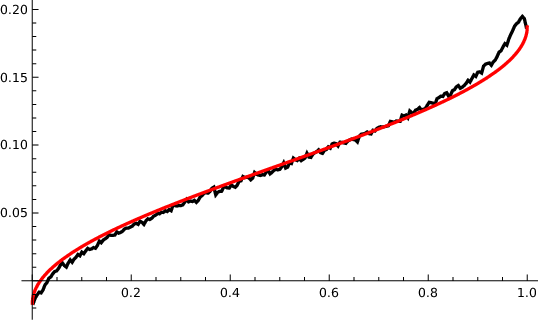} &
		\includegraphics[width=40mm]{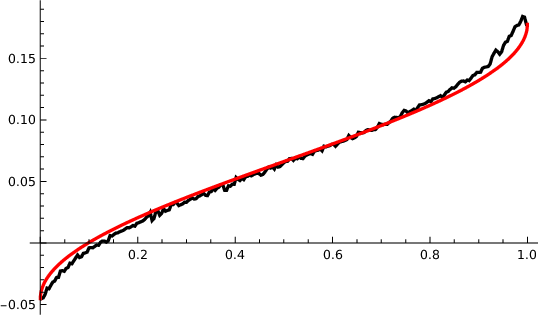} &
		\includegraphics[width=40mm]{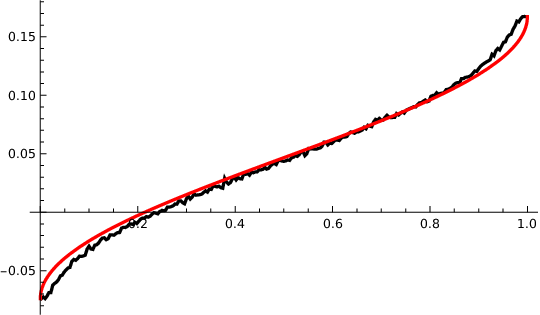} &
		\includegraphics[width=40mm]{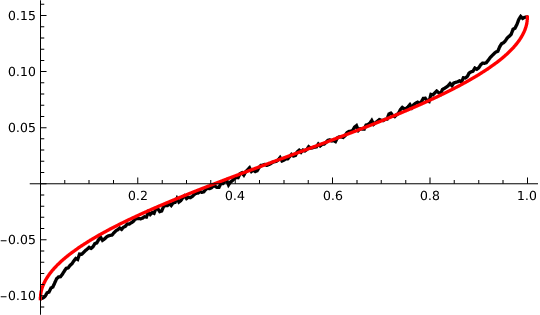}\\
		\includegraphics[width=40mm]{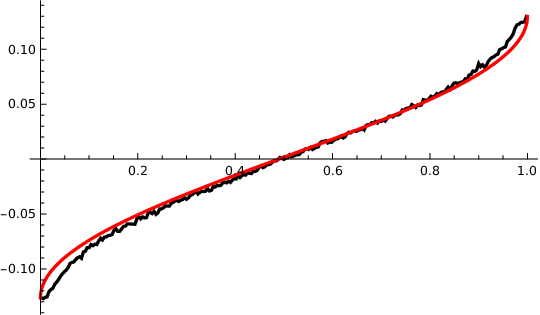} &
		\includegraphics[width=40mm]{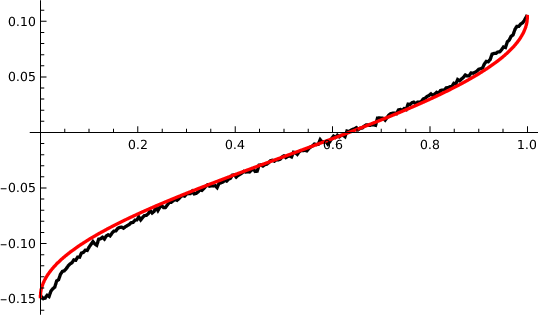} &
		\includegraphics[width=40mm]{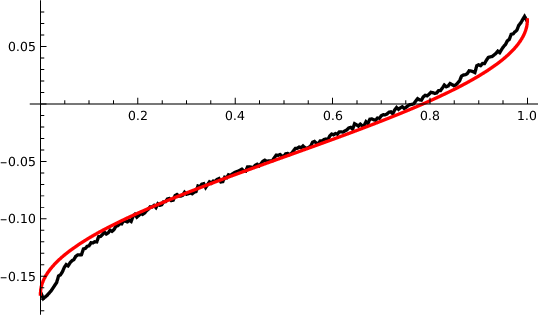} &
		\includegraphics[width=40mm]{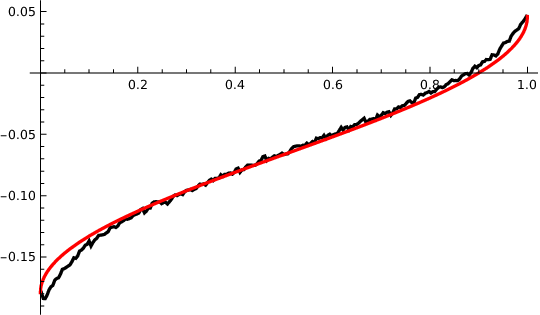} &
		\includegraphics[width=40mm]{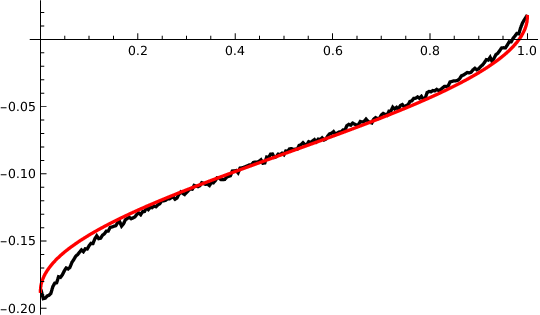}\\
		\includegraphics[width=40mm]{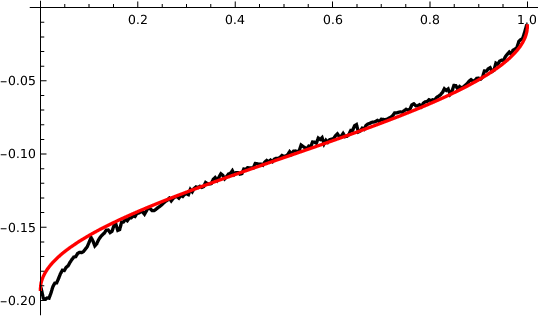} &
		\includegraphics[width=40mm]{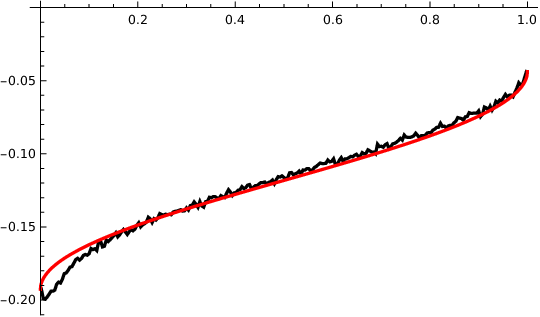} &
		\includegraphics[width=40mm]{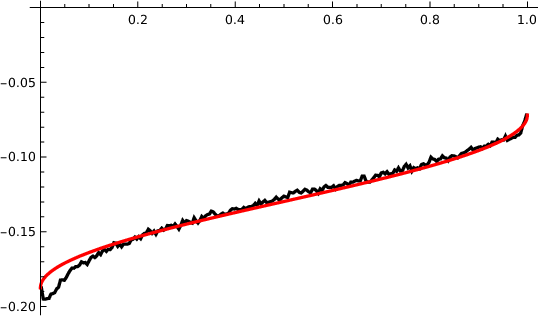} &
		\includegraphics[width=40mm]{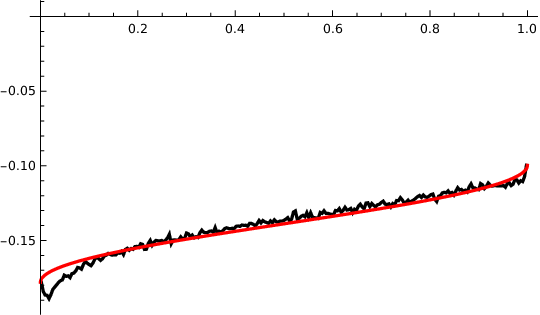} &
		\includegraphics[width=40mm]{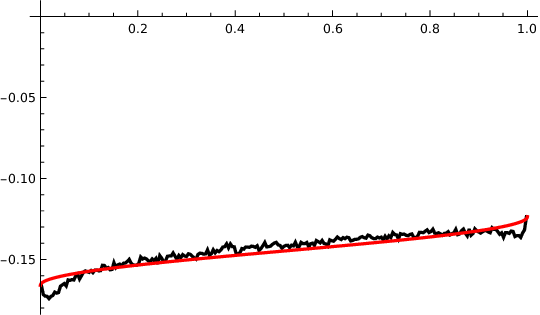}\\
		\includegraphics[width=40mm]{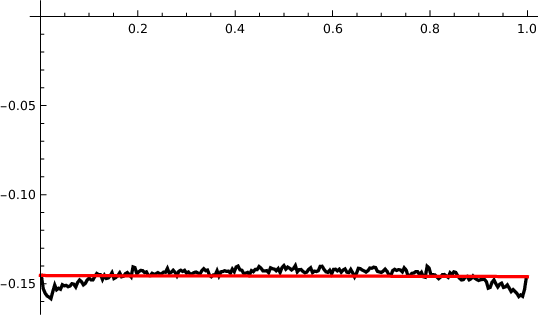} &
		\includegraphics[width=40mm]{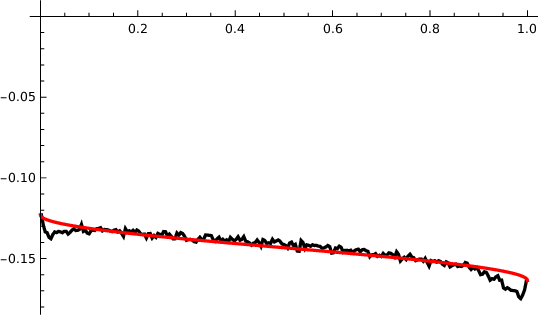} &
		\includegraphics[width=40mm]{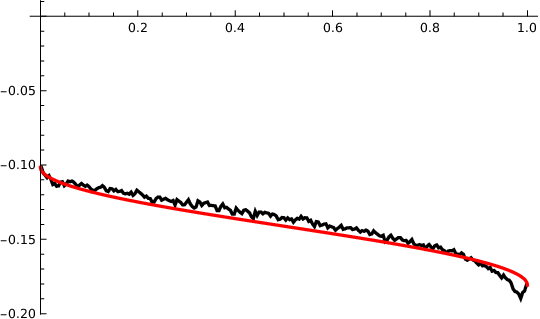} &
		\includegraphics[width=40mm]{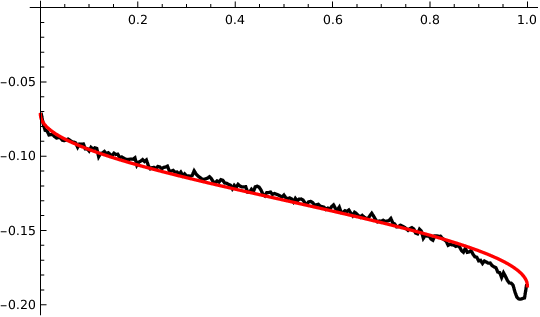} &
		\includegraphics[width=40mm]{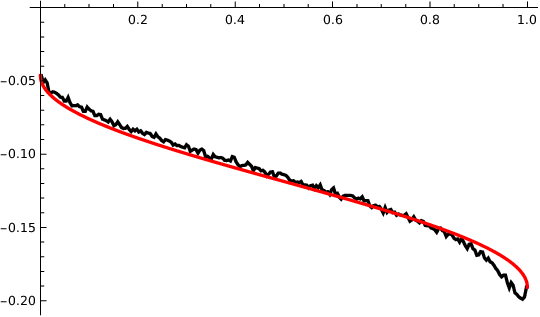}\\
		\includegraphics[width=40mm]{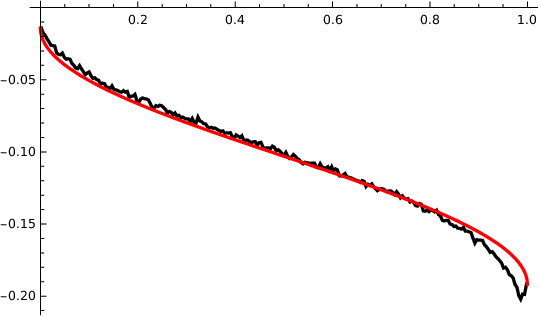} &
		\includegraphics[width=40mm]{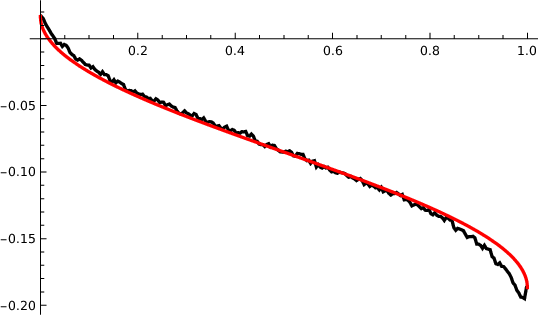} &
		\includegraphics[width=40mm]{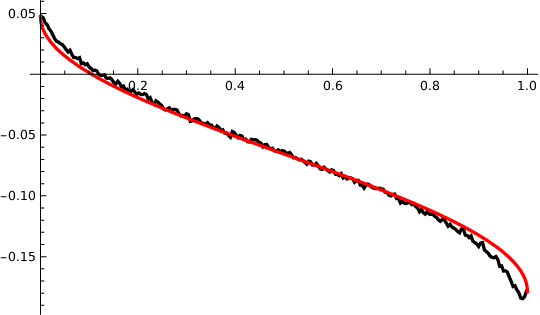} &
		\includegraphics[width=40mm]{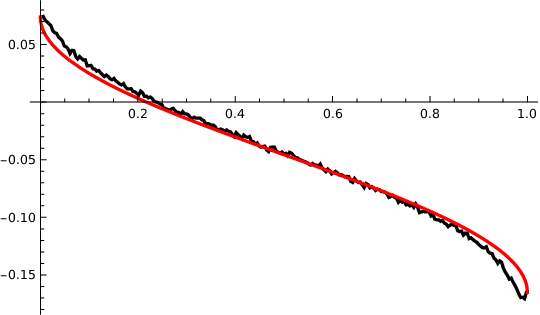} &
		\includegraphics[width=40mm]{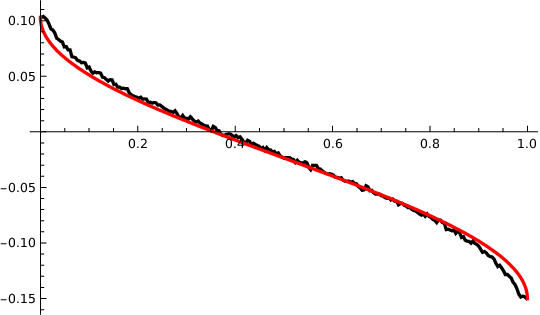}\\
		\includegraphics[width=40mm]{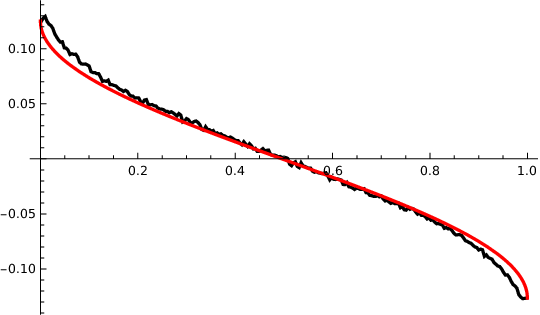} &
		\includegraphics[width=40mm]{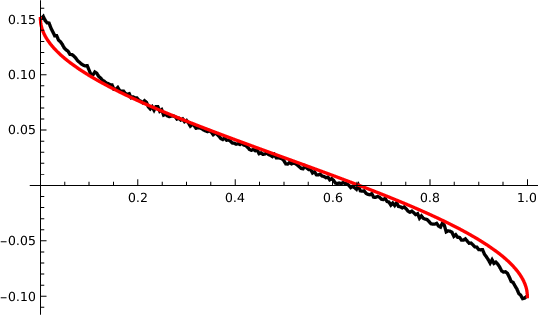} &
		\includegraphics[width=40mm]{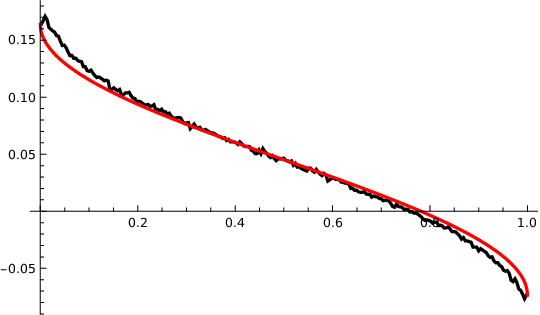} &
		\includegraphics[width=40mm]{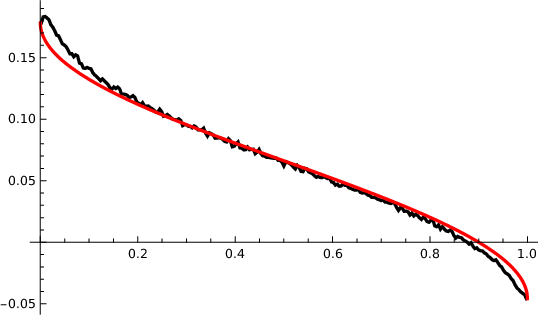} &
		\includegraphics[width=40mm]{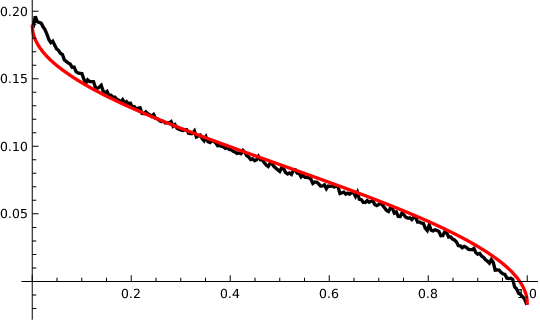}
	\end{tabular}
	\caption{Stationary magnetization profile plotted as a function of the position $x$. The black curve represents $\langle S_{i=xL}^{\rm z}\rangle/\Var(M)^{1/2}$ obtained from numerical simulations of the KPLL model with $L=255$ spins, $T=5000$ time steps, averaged over $10^{8}$ realizations, with boundary parameters $s_{a}+\ii s_{b}=0.1\,\ee^{\ii\theta}$ and $\theta$ taking the values $0,\pi/20,2\pi/20,\ldots$ from top to bottom and left to right. The solid red line is the small $\sigma_{a},\sigma_{b}$ asymptotics (\ref{magnetization profile small}).}
	\label{fig C1(x) open small}
\end{figure}

Figure~\ref{fig error magn profile} represents the difference between the KPLL simulations and the exact KPZ result for the stationary magnetization profile shown in figure~2 of the letter, with $s_{a}=1$, $s_{b}=-0.5$. Figures~\ref{fig C1(x) open}, \ref{fig int C1(x) open}, \ref{fig int C1(x) open diff} and \ref{fig C1(x) open small} represent the average stationary magnetization profile for the KPLL model for various values of the boundary magnetizations $s_{a}$ and $s_{b}$.

\clearpage
\section{Two-point correlation function}
\begin{figure}[b]
	\includegraphics[width=86mm]{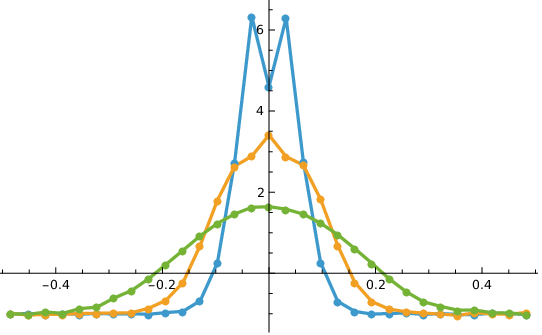}
	\includegraphics[width=86mm]{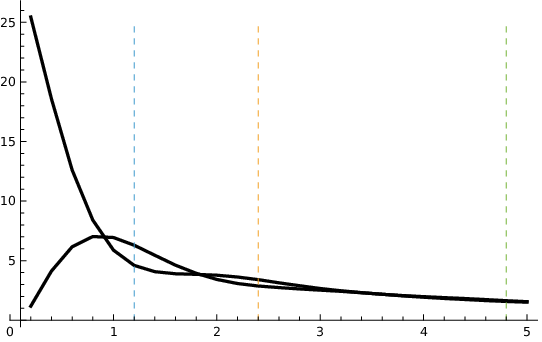}
	\begin{picture}(0,0)
		\put(-75,30){$i=0$}
		\put(-79,7){$i=1$}
	\end{picture}
	\caption{Stationary two-point function $C_{2}(x=i/L,T)$ for the Ishimori chain with $L=31$ spins and periodic boundaries, plotted as a function of the position $x$ for a few times $T$ (left) and as a function of $T$ for $i=0$ and $i=1$ (right ; the dashed lines materialize the times plotted on the left), showing a pronounced microscopic structure at earlier times around $i=0$, which prevents fixing the KPZ time $t$ reliably from $C_{2}(0,t)$, and is ultimately responsible for the bump observed in the Fourier modes $a_{k}(t)$ of $C_{2}(x,t)$, see figures~\ref{fig KPLL Fourier P} and \ref{fig Ishimori Fourier P}.}
	\label{C2 discrete}
\end{figure}

\begin{figure}
	\includegraphics[width=160mm]{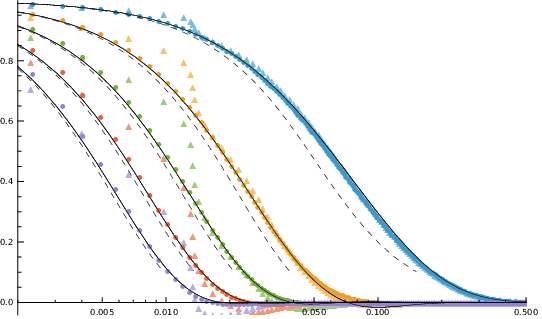}
	\begin{picture}(0,0)
		\put(-5,8){$t$}
		\put(-40,65){
			\put(0,3){\includegraphics[width=3mm]{KPLL_L32.eps}}\put(3,5){$L=32$}
			\put(0,-2.1){\includegraphics[width=3mm]{KPLL_L128.eps}}\put(3,-0.3){$L=128$}
		}
	\end{picture}
	\caption{First five Fourier coefficients $a_{1}(t),\ldots,a_{5}(t)$ (from top to bottom) of the stationary two-point function $C_{2}(x,t)\simeq c_{2}(x,t)$ with periodic boundaries, plotted as a function of time. The solid lines represent the exact results for the $a_{p}(t)$, and the dashed lines the short time approximation $\hat{c}_{2}^{\rm line}(2\pi k/\ell)$. The symbols represent the data from numerical simulations of the KPLL model with parameter $\tau=0.1$ averaged over $\approx10^{8}$ samples.}
	\label{fig KPLL Fourier P}
\end{figure}

\begin{figure}
	\includegraphics[width=160mm]{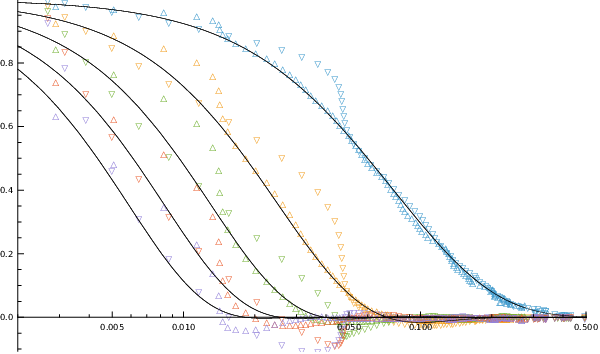}
	\begin{picture}(0,0)
		\put(-5,13){$t$}
		\put(-40,65){
			\put(0,3){\includegraphics[width=3mm]{Ishimori_L15.eps}}\put(3,5){$L=15$}
			\put(0,-2.1){\includegraphics[width=3mm]{Ishimori_L31.eps}}\put(3,-0.3){$L=31$}
		}
	\end{picture}
	\caption{First five Fourier coefficients $a_{1}(t),\ldots,a_{5}(t)$ (from top to bottom) of the stationary two-point function $C_{2}(x,t)\simeq c_{2}(x,t)$ with periodic boundaries, plotted as a function of time. The solid lines represent the exact results for the $a_{p}(t)$. The triangles represent the data from numerical simulations of the Ishimori chain averaged over $10^{5}$ samples.}
	\label{fig Ishimori Fourier P}
\end{figure}

Figure~\ref{C2 discrete} shows the discrete spatial structure responsible for the bump observed in the letter at shorter time $T$ and with a small number $L$ of spins for the two-point correlation function of the magnetization with periodic boundaries $C_{2}(x,t)\simeq c_{2}(x,t)$. Figures~\ref{fig KPLL Fourier P} and \ref{fig Ishimori Fourier P} show the Fourier coefficients in a larger format than in the letter, with the KPLL and Ishimori results split for better readability, and with averaging over more samples in the KPLL case.

\clearpage

